\documentclass[prb,twocolumn,showpacs]{revtex4}

\usepackage{times}
\usepackage{amsmath}
\usepackage{amsthm}
\usepackage{amssymb}
\usepackage{amsbsy}
\usepackage{graphicx}
\usepackage{pstricks}
\usepackage{color}
\usepackage{epstopdf}    
\usepackage{enumerate}
\usepackage{multirow} 

\begin{document}

\newcommand{\cred}{\color{red}}

\newcommand{\hexa}{\;
\pspicture(0,0.1)(0.35,0.6)
\psset{unit=0.75cm}
\pspolygon(0,0.15)(0,0.45)(0.2598,0.6)(0.5196,0.45)(0.5196,0.15)(0.2598,0)
\psset{linewidth=0.12,linestyle=solid}
\psline(0,0.15)(0,0.45)
\psline(0.2598,0.6)(0.5196,0.45)
\psline(0.5196,0.15)(0.2598,0)
\endpspicture\;}

\newcommand{\hexb}{\;
\pspicture(0,0.1)(0.35,0.6)
\psset{unit=0.75cm}
\psset {linewidth=0.03,linestyle=solid}
\pspolygon[](0,0.15)(0,0.45)(0.2598,0.6)(0.5196,0.45)(0.5196,0.15)(0.2598,0)
\psset{linewidth=0.12,linestyle=solid}
\psline(0.2598,0.6)(0,0.45)
\psline(0.5196,0.12)(0.5196,0.45)
\psline(0.2598,0)(0,0.15)
\endpspicture\;}

\newcommand{\smallhex}{ \;
\pspicture(0,0.1)(0.2,0.3)
\psset{linewidth=0.03,linestyle=solid}
\pspolygon[](0,0.0775)(0,0.225)(0.124,0.3)(0.255,0.225)(0.255,0.0775)(0.124,0)
\endpspicture
\;}


\begin{abstract}
Recently, quantum dimer models
have attracted a great deal of interest as a paradigm for the study of exotic quantum phases.
Much of this excitement has centred on the claim that a certain class of quantum dimer model 
can support a quantum $U(1)$~liquid phase with deconfined fractional excitations
in three dimensions.  
These fractional monomer excitations are quantum analogues of the magnetic monopoles 
found in spin ice.   
In this article we use extensive quantum Monte Carlo simulations to establish the ground-state 
phase diagram of the quantum dimer model on the three-dimensional 
diamond lattice as a function of the ratio $\mu$ of the potential to kinetic energy terms in the Hamiltonian.
We find that, for \mbox{$\mu_c = 0.75 \pm 0.02$}, the model undergoes a first-order quantum phase 
transition from an ordered ``{\sf R}-state'' into an extended quantum $U(1)$~liquid 
phase, which terminates in a quantum critical ``RK point'' for $\mu=1$.   
This confirms the published field-theoretical scenario. 
We present detailed evidence for the existence of the $U(1)$~liquid phase, and indirect evidence
for the existence of its photon and monopole excitations.  
Simulations are benchmarked against a variety of exact and perturbative results, and a 
comparison is made of different variational wave functions. 
We also explore the ergodicity of the quantum dimer model on a diamond lattice within a given 
flux sector, identifying a new conserved quantity related to transition graphs of dimer configurations.
These results complete and extend the analysis previously published 
in [O.~Sikora {\it et al.} Phys. Rev. Lett. {\bf 103}, 247001 (2009)].
\end{abstract}


\title{Extended quantum $U(1)$-liquid phase in a three-dimensional quantum dimer model}


\author{Olga Sikora}
\address{H. H. Wills Physics Laboratory, University of Bristol, Tyndall Avenue, Bristol BS8 1TL, UK.}

\author{Nic Shannon}
\address{H. H. Wills Physics Laboratory, University of Bristol, Tyndall Avenue, Bristol BS8 1TL, UK.}

\author{Frank Pollmann}
\address{Max-Planck-Institut f{\"u}r Physik komplexer Systeme, 01187 Dresden, Germany}

\author{Karlo Penc}
\address{Research Institute for Solid State Physics and Optics, H-1525 Budapest, P.O.B. 49, Hungary.}

\author{Peter Fulde}
\address{Max-Planck-Institut f{\"u}r Physik komplexer Systeme, 01187 Dresden, Germany}
\address{Asia Pacific Center for Theoretical Physics, Pohang, Korea}




\maketitle


\section{Introduction}


Dimer models, which describe the myriad
possible configurations of hard-core objects on bonds, have long been a 
touch-stone of statistical mechanics~\cite{fisher61, kastelyn61}.  
Quantum dimer models (QDM's), in which dimers are allowed to resonate between different 
degenerate configurations, were first introduced 
to describe antiferromagnetic correlations in high-$T_c$ superconductors~\cite{rokhsar88}.    
It has since been realized that QDM's arise naturally as effective models of many different 
condensed matter systems, and provide a concrete realizations of several classes of lattice
gauge theories.    
As such, they have become central to the theoretical search for new quantum phases and excitations.
A key question in this context is when, if ever, a QDM can support a liquid ground state?


\begin{figure}[th]
\begin{center}
\includegraphics[width=8cm]{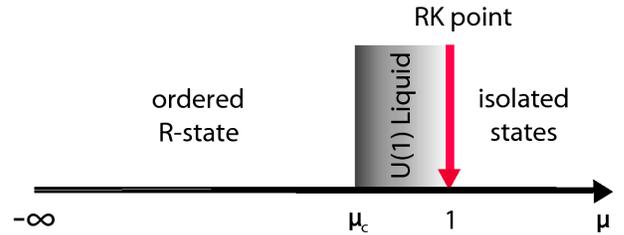}
\end{center}
\caption{
(Color online) 
Phase diagram of the quantum dimer model on a diamond lattice, as a function of the ratio $\mu$ 
of potential to kinetic energy, following~\protect\cite{moessner03,bergman06-PRB,sikora09}.
} 
\label{fig:phase_diagram}
\end{figure}


The answer to this question depends on lattice dimension and topology.  
In two dimensions (2D), the situation is relatively well-understood~\cite{diep04}, and 
a liquid ground state is known to exist in the QDM on the {\it non-bipartite} 
triangular~\cite{moessner01} and kagome~\cite{misguich02} lattices.
This liquid is gapped and has deconfined fractional excitations, which are vortices 
of an underlying $\mathbb{Z}_2$ (Ising) gauge theory.
Meanwhile,  for 2D {\it bipartite} lattices, the underlying gauge theory 
has a $U(1)$ character, akin to electromagnetism~\cite{huse03, moessner03}.  
In this case a liquid state is found only at a single, critical, ``Rokhsar-Kivelson'' (RK) 
point~\cite{rokhsar88}.  
Away from this, the system crystallizes into phases with broken lattice symmetries.  
For the QDM on a square lattice, these are columnar dimer and resonating plaquette 
phases~\cite{leung96,trousselet08}.  
%
A very similar phase diagram is found for the QDM on a honeycomb 
lattice~\cite{moessner01-honeycomb}.
%


Much less is known about QDM's in three dimensions (3D).   
However here field theoretical arguments suggest that a new, extended $U(1)$-liquid phase 
with gapless photon-like excitations might ``grow'' out of the RK point for a 3D QDM 
on a {\it bipartite} lattice~\cite{moessner03, bergman06-PRB} (see Fig.~\ref{fig:phase_diagram}). 
Similar claims have been made for the closely related quantum loop model in three 
dimensions~\cite{hermele04}.


In a recent Letter~\cite{sikora09}, we presented numerical evidence for the existence of an extended 
quantum $U(1)$~liquid phase with deconfined fractional excitations in a three-dimensional QDM, taken 
directly from its microscopic Hamiltonian. 
We considered the specific example of a QDM on a diamond lattice, which may serve as an effective model 
for the frustration found in various spinel compounds~\cite{bergman06-PRB,fulde02}. 
From zero-temperature GFMC simulations of clusters of up to $2000$ sites 
we were able to demonstrate the existence of a quantum $U(1)$-liquid phase 
for a range of parameters  $0.77 \pm 0.02 < \mu < 1$ bordering on the RK point [cf. Fig.~\ref{fig:phase_diagram}].  
These results complement a recent finite-temperature quantum Monte Carlo study of a microscopic 
model of interacting bosons which can be mapped onto an effective quantum loop model~\cite{banerjee08}.


In this article we document the evidence for these claims, taken from explicit calculations of
(i) the order parameter of the competing (dimer)-ordered phase [cf. Fig.~\ref{fig:lattice}], 
(ii) the ``string tension'' associated with fractional monomer excitations
and (iii) the finite-size scaling of a characteristic set of energy gaps associated with the 
U(1) liquid phase.
We also explore the internal consistency of these calculations,  cross-checking simulations 
against exact diagonalization of small clusters and perturbative expansions about exactly soluble 
limits of the model.


We pay particular attention to the thorny question of the ergodicity of GFMC
simulations in the presence of hidden quantum numbers. 
In the 2D QDM's studied previously, the quantum numbers are well understood, and 
it is clear which dimer configurations are connected by the QDM Hamiltonian.
This is not, however, the case for the 3D QDM considered in this article. 
Despite the fact that we can define a ``magnetic'' flux $\vec{\phi}$ which is a good 
quantum number --- in a manner similar to the square-lattice QDM in 2D --- the QDM 
Hamiltonian on a diamond lattice connects only sub-sets of dimer configurations within 
of each flux sector. 
This means that the QDM on diamond lattice is {\it not} ergodic within
a given flux sector, and great care must be taken to ensure that simulations 
accurately represent the true ground state for each value of flux $\vec{\phi}$.  
We shed some light on this issue by identifying a new, hidden, quantum 
number conserved by the QDM on diamond lattice.


The paper is structured as follows~:


In Section~\ref{model} we introduce the model and summarize the properties of the 
proposed $U(1)$-liquid phase. 
We demonstrate a way of constructing dimer configurations with different values of magnetic 
flux, and define an order parameter for the ordered ``{\sf R}-state'' found for negative values 
of $\mu$.   
%
In Section~\ref{simulation} we use Green's function Monte Carlo simulations to obtain explicit predictions 
for the $\mu$ dependence of the order parameter, monomer string tension and ground
state energy as a function of magnetic flux for a range of different cluster sizes and geometries.
We confirm the conjectured form of the phase diagram Fig.~\ref{fig:phase_diagram}, 
and use careful finite size scaling to obtain an improved estimate of the location of the phase 
transition from ordered to liquid phases in the thermodynamic limit as $\mu_c =  0.75 \pm 0.02$.  
In Section~\ref{ergodicity}, we explore the ergodicity of our simulations within a given flux sector 
and confirm that the techniques used give reliable estimates of ground state properties.  
We also identify an additional $\mathbb{Z}_2$ quantum number which divides each flux 
sector into two distinct parts.  
In Section~\ref{conclusions} we summarise our results and briefly discuss remaining 
open issues. 


The paper concludes with some technical appendices.   
In Appendix~\ref{fermi} we discuss the generalization to a fermionic QDM and show 
that it is equivalent to the bosonic case in lowest order. 
In Appendix~\ref{landau} we exhibit a Landau theory for the ordered, {\sf R}-state. 
In Appendix~\ref{VMC} we compare different variational wave functions, and use
these to investigate the transition from ordered to liquid ground states.
In Appendix~\ref{perturbation} we benchmark our Green's function Monte Carlo 
simulations against perturbation theory about exactly soluble limits of the model.   


\section{The Quantum Dimer Model and what you need to know about it}
\label{model}


\subsection{The model and its derivation}

\begin{figure}[h]
\begin{center}
\includegraphics[width=5cm]{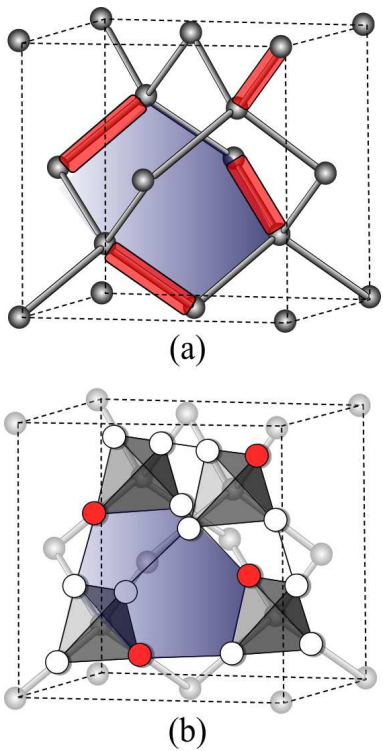}
\end{center}
\caption{
(Color online) 
(a) Cubic unit cell of the diamond lattice, with bonds occupied by dimers 
shown by thick red links.
The dimer configuration shown is the ordered ``{\sf R}-state'' found 
for negative $\mu$.
This contains a ``flippable'' hexagonal plaquette, which is shaded blue.
(b) Equivalent cubic unit cell of the pyrochlore lattice, with sites occupied
by hard-core bosons shown in red.
The boson configuration shown is the same ordered ``{\sf R}-state''.
} 
\label{fig:lattice}
\end{figure}

In the spirit of Ref.~\onlinecite{rokhsar88}, we study the quantum dimer model (QDM)   
\begin{eqnarray}
{\mathcal H}_{\sf QDM} =  -g  {\mathcal K}_{\sf f} + \mu {\mathcal N}_{\sf f} 
\label{eq:QDM}
\end{eqnarray}
where the matrix elements of 
\begin{eqnarray}
{\mathcal K}_{\sf f}
&=& 
\sum_{\smallhex}\left( \Big|\hexa\Big\rangle \Big\langle\hexb\Big| 
+  \Big|\hexb\Big\rangle  \Big\langle\hexa\Big|\right)
\label{eq:K}
\end{eqnarray}
connect different, degenerate dimer configurations by cyclically permuting 
dimers on six-bond ``flippable'' plaquettes of alternating filled and empty 
links within the diamond lattice [cf. Fig.~\ref{fig:lattice}], 
while 
\begin{eqnarray}
{\mathcal N}_f 
&=& \sum_ {\smallhex}\left(
\Big|\hexa\Big\rangle\Big\langle\hexa\Big|
+\Big|\hexb\Big\rangle\Big\langle\hexb\Big| \right)
\label{eq:Nf}
\end{eqnarray}
counts the number of such flippable plaquettes.
In what follows we will consider $g > 0$, with $-\infty < \mu \le g$.  


This QDM Hamiltonian acts on the $\approx 1.3^{N/2}$ different dimer configurations which cover the links 
the $N$-bond diamond lattice~\cite{nagle66}.
This Hilbert space breaks up into different subsectors $\lambda_{\sf c}$, defined by the sets of 
dimer configurations which are connected by the matrix elements of ${\mathcal K}_{\sf f}$. 
The number and nature of these subsectors remains an open problem, which will be discussed
at length in Section~\ref{ergodicity} of this article.


Crucially, however, {\it all} off-diagonal matrix elements of Eq.~(\ref{eq:QDM}) are zero or negative.
By the Frobenius-Perron theorem, the lowest energy state of the QDM within {\it any} given 
subsector of the Hilbert space $\lambda_{\sf c}$ must then be a nodeless superposition 
of dimer configurations 
\begin{eqnarray}
\mid \psi_0 \rangle_{\lambda_{\sf c}}
= \sum_{{\sf c} \in \lambda_{\sf c}} a_{\sf c} |{\sf c} \rangle
\end{eqnarray}
with real, {\it positive} coefficients $a_{\sf c}$.   
As a result, Quantum Monte Carlo simulations of Eq.~(\ref{eq:QDM}) are {\it not} 
subject to a sign problem.   


Exactly at the RK point $\mu = g$, the QDM Hamiltonian can be written as a 
sum of projectors
\begin{eqnarray}
{\mathcal H}_{RK} 
 &=& g \sum_{\smallhex} 
\left[ \Big|\hexa\Big\rangle - \Big|\hexb\Big\rangle \right]
\left[ \Big\langle\hexa\Big| - \Big\langle\hexb\Big| \right] 
\label{eq:RK}
\end{eqnarray}
In this case the ground state in {\it any} subsector of the Hilbert space $\lambda_{\sf c}$ 
is the zero-energy eigenstate given by the equally weighted sum over {\it all} dimer configurations 
within that subsector.   
\begin{eqnarray}
| {\sf RK} \rangle_{\lambda_{\sf c}} \propto \sum_{{\sf c} \in \lambda_{\sf c}}  |{\sf c} \rangle
\end{eqnarray}
%


The QDM Eq.~(\ref{eq:QDM}) can be derived from various microscopic models. 
One example is a hard-core boson model on a pyrochlore lattice 
formed of corner sharing tetrahedra
\begin{eqnarray}
{\mathcal H}_{\sf tV} = -t \sum_{\langle ij \rangle} (b_i^{\dagger}b_j  + H.c.) 
+ V\sum_{\langle ij \rangle} n_i n_j  
\label{eq:tVmodel}
\end{eqnarray}
at quarter filling, in the limit of large nearest-neighbor interactions $V \gg t$.
In this case, the low energy configurations are those without bosons on neighbouring sites. 
This imposes the constraint of placing {\it exactly} one boson in each tetrahedron.


The diamond lattice is the medial lattice of the pyrochlore lattice, formed by connecting
the centres of all tetrahedra (see Fig.~\ref{fig:lattice}), and 
and these constrained states are in exact one-to-one correspondence 
with the dimer coverings of a diamond lattice.
The lowest-order process in degenerate perturbation theory which connects 
these states is a cyclic exchange of bosons around the smallest, hexagonal plaquette
of the pyrochlore lattice.
This occurs at third order in the hopping of bosons, giving $g=12t^3/V^2$.  


Another physically motivated model which reduces to the quantum dimer model on the diamond lattice
is the easy-axis quantum antiferromagnet on a pyrochlore lattice 
\begin{eqnarray}
\label{eq:XXZmodel}
{\mathcal H}_{\sf XXZ} = J_{xy} \sum_{\langle ij \rangle} (S^x_iS^x_j + S^y_iS^y_j) 
+ J_z \sum_{\langle ij \rangle} S^z_iS^z_j \nonumber\\
- h \sum_i S^z_i  
\end{eqnarray}
at half-magnetization $\langle S^z \rangle \equiv S/2$.  
For $S=1/2$ this is exactly equivalent to the hard-core boson problem considered above, with each 
occupied site corresponding to a ``up'' spin, and each empty site to a ``down'' spin, 
with the caveat that the kinetic energy term $J_{xy}$ now has the opposite sign, 
and so $g = -12J_{xy}^3/J_z^2$. 

\begin{figure}[h]
\begin{center}
\includegraphics[width=5cm]{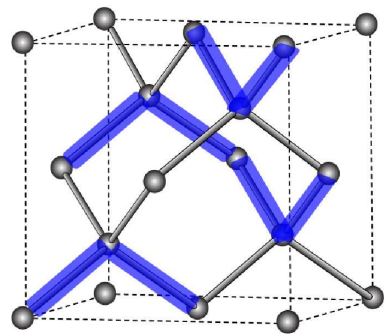}
\end{center}
\caption{
(Color online) 
The subset of diamond lattice bonds $\Lambda$ (thick, blue, shaded bonds) 
used in the unitary transformation Eq.~(\ref{eq:unitary-transformation}) 
which maps $g \to -g$ in Eq.~(\ref{eq:QDM}).
} 
\label{fig:unitary-transformation}
\end{figure}

At first sight this might seem to imply that the $XXZ$ model Eq.~(\ref{eq:XXZmodel}) with 
$g < 0$ has a different ground state from the a $t$-$V$ model with $g >0$ Eq.~(\ref{eq:tVmodel}). 
However the global sign of $g$ can be changed by the simple unitary transformation
\begin{eqnarray}
|c\rangle \rightarrow \exp [i \pi {\mathcal{N}_{\Lambda}}/2] |c\rangle
\label{eq:unitary-transformation}
\end{eqnarray}
where $|c\rangle$ is an arbitrary dimer configuration and $\mathcal{N}_{\Lambda}$
counts the number of dimers which occupy the subset of diamond lattice bonds
$\Lambda$ shown in Fig.~\ref{fig:unitary-transformation}.
The cyclic permutation of dimers on any six-bond ÒflippableÓ plaquette changes 
$\mathcal{N}_{\Lambda}$ by two, and so the unitary transformation 
Eq.~(\ref{eq:unitary-transformation}) maps $g \to -g$. 
Moreover,  exactly the same quantum dimer model Eq.~(\ref{eq:QDM}) can
be derived from a fermionic $t$-$V$ model, i.e. Eq.~(\ref{eq:tVmodel}) with 
Fermi operators substituted for the bosonic ones. 
This mapping is described in Appendix~\ref{fermi}. 
%


An effective Hamiltonian of the QDM form Eq.~(\ref{eq:QDM}) can also be derived from 
Eq.~(\ref{eq:XXZmodel}) for larger values of spin.
For $S=3/2$, where the model might be of relevance to Cr spinels, this was accomplished 
by Bergman {\it et al.}~\cite{bergman06-PRL}, who found $\mu/g \approx  -7$.
A two-dimensional analogue of Eq.~(\ref{eq:QDM}) might also be realized using 
cold atoms~\cite{ruostekoski09}.
Finally, quantum dimer models also arise as effective models of spin--orbital 
models~\cite{vernay06}.


Since the goal of this paper is to determine the ground state phases of Eq.~(\ref{eq:QDM}), 
and not to relate it to any particular physical system, in what follows we set $g=1$, and study 
the model as a function of the single adjustable parameter $\mu$.  
For \mbox{$\mu\rightarrow-\infty$} the potential energy dominates, and the
ground state of Eq.~(\ref{eq:QDM}) is the set of dimer
configurations which maximise the number of ``flippable'' hexagons, with 
one out of four hexagons being flippable.   
This is the so-called \mbox{{\sf R}-state} \cite{bergman06-PRL}, illustrated in 
Fig.~\ref{fig:lattice}, which has a cubic unit cell and is eight-fold degenerate.


At the RK point $\mu=1$, the ground state is the equally weighted sum of all 
possible dimer configurations~\cite{rokhsar88}, while for $\mu>1$, ground states 
are the many ``isolated states'' which contain {\it no} flippable hexagons, and 
so have no kinetic energy.  
The interest of the QDM therefore rests in the parameter range 
\mbox{$-1 \lesssim \mu \leq 1$}, for which kinetic and potential energy compete 
on an equal footing.   
It is this range of parameters which we consider below.


\subsection{Effective electrodynamics}
\label{electrodynamics}


For purpose of comparison with numerics, we now briefly review the field theory arguments 
underlying the proposed $U(1)$~liquid phase. 
The defining property of a dimer model is that every lattice site is touched by {\it exactly} one dimer.   
The diamond lattice is bipartite --- we can divide it in two sublattices, $A$ and $B$, such
that every every link (bond) of the lattice connects an $A$ sublattice site with a $B$ sublattice site.  
We can use this property to associate a magnetic flux vector ${\bf B}$ with each bond of the lattice.
Bonds not occupied by a dimer are assigned {\it one} (arbitrary) unit of magnetic flux,  directed 
from $B$ to $A$.
Where a dimer is present, we assign {\it three} units of flux to that bond, directed from $A$ to $B$
\footnote{We note that these, arbitrary, units differ from those defined in Eq.~( \ref{eq:phi_vs_n}) by a factor $4/\sqrt{3}$.}.
 

In this language, the constraint that every lattice site is touched by one dimer becomes the condition 
\begin{eqnarray}
\nabla \cdot \mathbf B = 0.
\end{eqnarray}
We resolve this constraint by writing 
\begin{eqnarray}
\mathbf B=\mathbf{\nabla} \times \mathbf A,
\end{eqnarray}
where the gauge field $\mathbf A$ is defined on the {\it sites} of the original diamond
lattice while the fictitious magnetic field $\mathbf B$ is defined on its {\it bonds}.
We chose to work in the Coulomb gauge $\nabla \cdot \mathbf A = 0$.  
In the absence of quantum effects, the statistical description of the dimer model 
reduces to a relatively straightforward problem in magnetostatics, and the ground state of 
the dimer model is found to be classical $U(1)$-liquid.
Correlations between dimers have a dipolar form in real space, and the constraint
$\nabla \cdot \mathbf B = 0$ manifests itself as a set of ``pinch points'' in the structure factor
in k-space~\cite{youngblood80, huse03, henley05, isakov06, henley10}.   
Exactly this situation is realized in the rare-earth magnet spin ice~\cite{fennel09}.


The purpose of this paper is to treat the quantum effects which arise from the 
tunnelling of the system from one dimer configuration to another, as described by Eq.~(\ref{eq:QDM}).
This tunnelling introduces fluctuations in time of the gauge field $\mathbf A$, which 
in turn give rise to an effective electric field 
\begin{eqnarray}
\mathbf E= - \partial_t\mathbf A.
\end{eqnarray}
%
%
None the less, the total magnetic flux $\phi = \int d {\mathbf S}\cdot{\mathbf B} $ 
through any plane in the lattice remains a conserved quantity, since the dynamics 
present in the quantum dimer model only permit a reversal in the sense of flux around 
a closed loop.


This representation clearly has a lot in common with conventional electromagnetism  
and, following Ref.~\onlinecite{moessner03},  we can use this analogy to write down a plausible 
long-wavelength action for the QDM on a diamond lattice 
\begin{eqnarray}
\mathcal{S}=\int d^3xdt\left[ \mathbf E^2 - c^2 \mathbf B^2 \right],
\label{eq:maxwellS}
\end{eqnarray}
where $c^2 > 0$. This is the Maxwell action of conventional electromagnetism, 
and the system must posses linearly dispersing ``photons'' (transverse excitations 
of the gauge field ${\bf A}$) with speed of light $c$.     
These conditions were argued to be realized in a quantum $U(1)$-liquid state 
bordering the RK point in three-dimensional quantum dimer models on a bipartite 
lattice for $\mu \lesssim 1$~[\onlinecite{moessner03}], leading to the conjectured 
phase diagram Fig.~\ref{fig:phase_diagram}.  
An exactly parallel story can be told about the quantum loop model 
in three dimensions~\cite{hermele04}.


These ``emergent'' photons are the signature feature of the proposed $U(1)$-liquid state, 
and offer a beautiful realization of Maxwell's laws in a condensed matter system.   
However, for the purposes of this study, we wish to emphasize that Eq.~(\ref{eq:maxwellS}) also 
contains information about the {\it finite size} scaling of the energy spectra. 
A flux $\phi$ through a cluster of volume $L^3$ corresponds to an average magnetic 
field $B = \phi/L^2$.  
In the $U(1)$-liquid state, this magnetic field is uniformly distributed on the ``coarse-grained'' 
scale of the effective action Eq.~(\ref{eq:maxwellS}).  
It then follows from Eq.~(\ref{eq:maxwellS}) that the energy difference 
$\Delta_\phi = E_\phi - E_0$ between the ground state of the zero-flux sector, and 
the lowest energy state of the sector with flux $\phi$ scales as
\begin{eqnarray}
\Delta_ \phi = E_\phi - E_0 = c^2 \frac{\phi^2}{L}\ .
\label{eq:gaps}
\end{eqnarray}
In the thermodynamic limit $\Delta_ \phi \to 0$ and different flux sectors should be treated as degenerate ground
states. 
A second finite-size prediction of Eq.~(\ref{eq:maxwellS}) is that the ground state energy of 
the $U(1)$-liquid state should have a finite size correction $\Delta E$ from the zero-point energy 
of photons which in leading order scales as 
\begin{eqnarray}
\frac{\Delta E (L)}{N} 
\sim 
\left[\int_{\frac{2\pi}{L}}^{\Lambda'} dk - \int_0^{\Lambda'} dk\right] k^2 c k \sim  - \frac{c}{L^4} 
\label{eq:zero_point}
\end{eqnarray}
where $\Lambda'$ is a size-independent momentum cutoff.
In this paper we make use of both Eq.~(\ref{eq:gaps}) and Eq.~(\ref{eq:zero_point})
to study finite size properties of  Eq.~(\ref{eq:QDM}).


\subsection{Clusters used in simulation}
\label{clusters}


Our evidence for the existence of a quantum $U(1)$-liquid phase is taken from simulation
of the quantum dimer model Eq.~(\ref{eq:QDM}) on finite-size clusters of the diamond lattice
with periodic boundary conditions.
For the size of a cluster use the notation in which $N$ is the number of pyrochlore lattice 
sites, corresponding to the diamond lattice bonds. 
The periodic boundary conditions permit us to associate a set of (integer) topological 
quantum numbers $\vec{\phi} = (\phi_1,\phi_2,\phi_3)$ with the flux through a set of 
orthogonal planes.
This reduces the Hilbert space of the problem from the $\approx 1.3^{N/2}$ different dimer configurations
to a set of discrete sectors with given $\vec{\phi}$.



An important lesson from the exact diagonalization of small clusters is that the matrix elements
of Eq.~(\ref{eq:QDM}) do {\it not} connect all dimer configurations within a given value of $\vec{\phi}$.  
None the less, the size of the largest blocks is still too large to permit the 
exact diagonalization of clusters with more that 128 diamond lattice bonds.
These clusters are very small by the standards of  the (implicitly) course-grained field 
theory Eq.~(\ref{eq:maxwellS}), and quantum Monte Carlo simulations must therefore
be used to determine the ground state of the model.


\begin{table}
\caption{
Cluster geometries used in simulations.    
The first column gives the short notation for cluster geometry used in the text; the second column, the number of 
diamond lattice bonds; the last three columns show the vectors ${\bf g}_i$ which define the translation vectors of the cluster.  
The integer $l$ defines the size of the cluster, within a given family.
The [100] clusters have side of length of $L=2 l$, and contain $N = 2 L^3$ bonds.
}  
\begin{ruledtabular}
\begin{tabular}{c|c|c|c|c}
\label{table_g}
 & number of bonds & ${\bf g}_1$ & ${\bf g}_2$ & ${\bf g}_3$ \\
\hline
$[$100$]$   &  $16\,l^3$   &   $ 2l\,(1,0,0) $  & $ 2l\,(0,1,0)  $ &  $ 2l\,(0,0,1) $ \\
$[$110$]$   &  $32\,l^3$   &   $ 2l\,(1,1,0)  $ & $ 2l\,(0,1,1)  $ &  $ 2l\,(1,0,1) $\\
$[$111$]$   &  $64\,l^3$   &   $ 2l\,(1,1,-1) $ & $ 2l\,(1,-1,1) $ &  $ 2l\,(-1,1,1)$ \\
\end{tabular}
\end{ruledtabular}
\end{table}


The quantum Monte Carlo method we chose is GFMC (Green's function Monte Carlo), which 
constructs Monte Carlo updates using only the Hamiltonian dynamics of Eq.~(\ref{eq:QDM}).
This has the advantage that all quantum numbers (including magnetic flux) are preserved
in simulation.
However since not all of these quantum numbers are known {\it a priori}, 
great care must be taken to simulate in the appropriate subsectors of the Hilbert space.
This is an issue which we discuss at length in Section~\ref{ergodicity}.
Moreover, because an exponential number of dimer configurations contribute to the ground 
state wave function of the $U(1)$-liquid, quantum Monte Carlo simulations are themselves 
limited to clusters of up to 2000 diamond lattice bonds.


In order to gain the most from finite size scaling, we therefore consider a range of different
cluster geometries.  
We restrict ourselves to clusters which are compatible with the 16-bond unit cell of 
the \mbox{{\sf R}-state}, and group them into families of clusters with the same shape. 
These are summarized in \mbox{Table~ \ref{table_g}}.  
The main family of clusters used in our simulations consists of clusters with $2L^3$ diamond lattice bonds 
(pyrochlore lattice sites) with side of length $L={4,6,...,16}$, which have the full (cubic) symmetry of the diamond lattice.
We denote this family as [100], since the edges of the cluster are parallel to cubic lattice axes.
The two other families of clusters considered are denoted [110] and [111]. 
All results are quoted for [100] clusters, unless specified otherwise.


\subsection{String defects and flux sectors}
\label{strings}


\begin{figure}[h]
\begin{center}
\includegraphics[width=6cm]{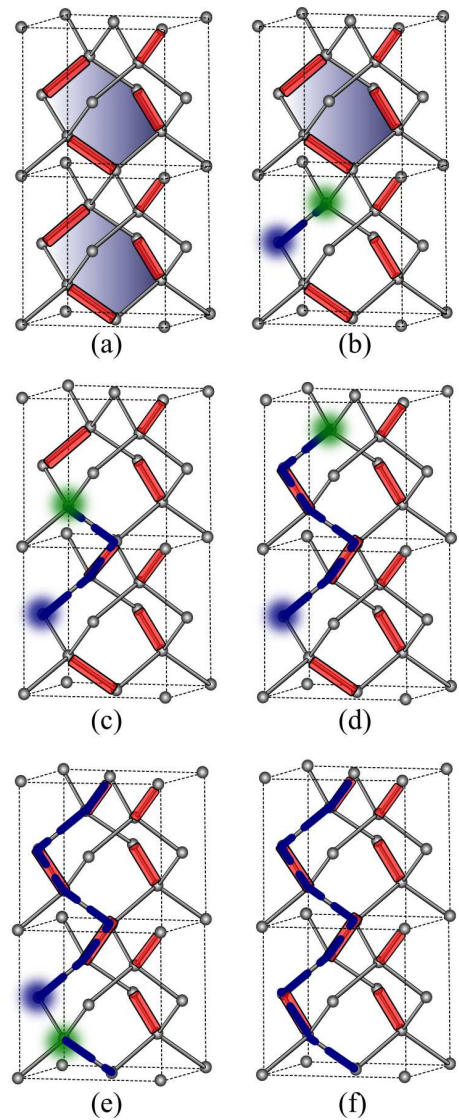}
\end{center}
\caption{
(Color online) 
Construction of the ``string'' defect (one magnetic flux) configuration from the maximally 
flippable {\sf R}-state configuration (assuming periodic boundary conditions): 
(a) two adjacent 16-bond unit cells of the diamond lattice within the {\sf R}-state, 
with flippable hexagons denoted by blue shading; 
(b)  a pair of monomer excitations created by removing a dimer; 
(c), (d) and (e) monomers are separated.   
The dashed line shows the ``string'' of displaced dimers which connects them.  
(f) one of the monomers has traversed the cluster and the dimer is re-introduced.
The resulting configuration has a single string excitation and a nonzero magnetic flux.
}
\label{fig:string}
\end{figure}


As discussed in Section~\ref{electrodynamics} and Section~\ref{clusters} above, any 
given dimer configuration on the diamond lattice can be classified according to its total 
``magnetic'' flux 
\begin{eqnarray}
\vec{\phi} = (\phi_1,\phi_2,\phi_3)
\end{eqnarray}
where each component of $\vec{\phi}$ is a (topological) quantum number 
conserved under the dynamics of the quantum dimer model Eq.~(\ref{eq:QDM}).
%


In a finite size cluster, each component of flux $\phi_i$ serves as a winding number 
which takes on an extensive number of (discrete) values.
These are bounded by the size of the system, and can be expressed as combinations 
of the four occupation numbers $(n_1,n_2,n_3,n_4)$ which count the number of dimers on 
each of the four fcc sublattices of diamond-lattice bonds (equivalently, pyrochlore lattice sites).
These occupation numbers are themselves invariant under the dynamics of the QDM.  
However the mapping between the occupation number representation 
(cf. Ref.~\onlinecite{bergman06-PRB}), and the flux representation of these conserved quantities 
depends on the specific cluster geometry.  


For the (100) cluster series we obtain the following expression: 
\begin{eqnarray}
\phi_x &=&  \frac{1}{2L}  (n_1-n_2-n_3+n_4) \nonumber\\  
\phi_y &=&  \frac{1}{2L}  (n_1-n_2+n_3-n_4) \nonumber\\  
\phi_z &=& \frac{1}{2L} (n_1+n_2-n_3-n_4),
\label{eq:phi_vs_n}
\end{eqnarray} 
where the corresponding positions of the sites in the 4-site elementary unit 
cell of the pyrochlore lattice are numbered 
\begin{eqnarray}
\mathbf{r}_1 &=& (1, 1, 1)/4  \nonumber\\ 
\mathbf{r}_2 &=& (-1,-1,1)/4\nonumber\\ 
\mathbf{r}_3 &=& (-1,1,-1)/4\nonumber\\ 
\mathbf{r}_4 &=& (1,-1,-1)/4\nonumber
\end{eqnarray} 
It follows that $-L^2/4 \le \phi_i \le L^2/4$ for a cluster with $N=2L^3$ bonds 
in the diamond lattice ({\it i.e.} the number of sites in the pyrochlore lattice). 
In the highly symmetric ${\sf R}$--state (cf.~Fig.~\ref{fig:lattice})
\mbox{$n_1 = n_2 = n_3 = n_4$}.  
This state therefore belongs to the zero-flux sector \mbox{$\vec{\phi} = (0,0,0)$}.   


In order to test the predictions of the effective field theory, notably Eq.~(\ref{eq:gaps}), 
we need to be able to construct other states in neighbouring flux sectors.   
The flux quantum numbers $(\phi_1,\phi_2,\phi_3)$ are conserved under {\it all} local operations 
which connect configurations satisfying the dimer constraint.  
However, we can construct states with finite magnetic flux, starting from the ${\sf R}$--state, by 
creating non-local ``string'' defects which traverse the periodic boundaries of the cluster.


We do this following the procedure illustrated in Fig.~\ref{fig:string}.  
When a dimer is removed from the lattice, two monomers (sites not touched by a dimer) are
created. 
Subsequently, we can separate these monomers, creating a ``string'' of dimers which have been 
translated by exactly one bond.
Moving a dimer by one bond reverses the sense of the magnetic field ${\mathbf B}$ 
associated with that dimer.   
We can therefore construct a state with a different flux by moving one of the monomers 
across the periodic boundary of the cluster in such a way that the two monomers meet 
and the dimer can be re-introduced.  
The resulting configuration fulfills again the zero-divergence condition on ${\mathbf B}$, but the 
magnetic flux through the periodic boundary traversed by the monomer has changed by exactly 
one unit. 
For a cubic cluster this is the smallest magnetic flux possible in the system, and we choose 
convention where $\phi=1$ in the new configuration.
States with higher flux can be constructed by repeating the procedure above.


The monomers created by removing a dimer are excitations with fractional quantum numbers
which depend on the original microscopic model, e.g. they might be fractional charges 
$e/2$~[\onlinecite{fulde02}], or ``spinons" carrying spin $s=1/2$ [\onlinecite{bergman06-PRL}].  
In field-theoretical terms, they are the magnetic monopoles of a $U(1)$ gauge theory, and the 
central question addressed in this paper is whether the QDM on a diamond lattice can support
a quantum $U(1)$ liquid phase in which these monopoles are deconfined.  
In a finite size system, deconfined monopoles must be free to traverse the periodic boundaries
of the cluster, and so the string defects described above also provide a direct test of monopole 
confinement --- a necessary condition for monopole deconfinement
in a periodic cluster is that the ground states in flux sectors connected by strings 
should be degenerate [c.f. Eq.~(\ref{eq:gaps})].   
We return to this idea below.  


\subsection{Competing ordered phase}


\begin{figure}[h]
\begin{center}
\includegraphics[width=8cm]{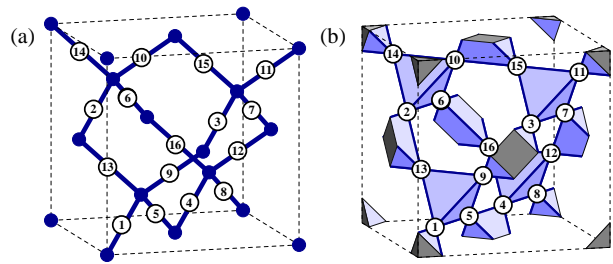}
\end{center}
\caption{
(Color online) 
Numbering convention in the definition of the order parameter $m_{\sf R}$
for the \mbox{{\sf R}-state} (Eq.~\ref{eq:op}): 
(left) cubic cell of pyrochlore lattice with 16 sites, 
(right) corresponding diamond lattice cell with 16 bonds.
}
\label{Fig:op_conv}
\end{figure}


\begin{table}
\caption{
Coefficients of bond occupation numbers within the six-dimensional irreducible representation
used to define an order parameter for the ordered \mbox{{\sf R}-state}, 
following Eq.~(\ref{eq:op}).  
}  
\begin{ruledtabular}
\begin{tabular}{c|c|c|c|c|c|c|c|c|c|c|c|c|c|c|c|c}
\label{table}
 & 1 & 2 & 3 & 4 & 5 & 6 & 7 & 8 & 9 & 10 & 11 & 12 & 13 & 14 & 15 & 16 \\
\hline
$f_1$ &1& -1& -1&  1& -1&  1& 1& -1& -1& 1& 1& -1& 1& -1& -1& 1 \\
$f_2$  &1&  1& -1& -1& -1& -1& 1& 1& 1& 1& -1& -1& -1& -1& 1& 1\\
$f_3$ &1& -1&  1& -1&  1& -1&  1& -1& -1& 1&  -1& 1& -1& 1& -1& 1\\
$f_4$ &1& -1& -1&  1& -1&  1&  1& -1& 1& -1& -1& 1& -1& 1& 1& -1\\
$f_5$ &1&  1& -1& -1&  1&  1&  -1& -1& -1& -1& 1& 1& -1& -1& 1& 1\\
$f_6$ &1& -1&  1& -1& -1&  1& -1& 1& -1& 1& -1& 1& 1& -1& 1& -1\\
\end{tabular}
\end{ruledtabular}
\end{table}


In order to determine the phase diagram of the quantum dimer model on a diamond lattice 
(cf. Fig.~\ref{fig:phase_diagram}), we need also to characterize the competing ordered 
\mbox{``{\sf R}-state''} [Refs.~\onlinecite{bergman06-PRL,bergman06-PRB}].   
The \mbox{{\sf R}-state} (illustrated in Fig.~\ref{fig:lattice}) 
occupies the 8-site (16-bond) cubic unit cell of the diamond lattice 
(16-site on the pyrochlore lattice), and is \mbox{8-fold} degenerate. 
The \mbox{{\sf R}-state} breaks the inversion symmetry of the diamond lattice, its 
8-fold degeneracy is best thought of as $2\times 4$ states with opposite chirality. 
The $4$ states with equal chirality can be related to each other with rotations or translations.  
In terms of the gauge field, this \mbox{8-fold} degeneracy can be thought of as a \mbox{4-fold} 
choice of $111$ axis (for a given tetrahedron), and a two fold choice of chirality about that axis.   
However for purposes of simulation it is easier to keep track of an order parameter
defined in terms of the occupation numbers for dimers on diamond lattice bonds.


We define this through a 6-dimensional irreducible representation of the diamond lattice 
point group as follows 
\begin{eqnarray}
m_{\sf R} = \sqrt{\sum_{\eta=1}^{6}m^2_{{\sf R},\eta}},
\label{eq:op}
\end{eqnarray}
where 
\begin{eqnarray}
m_{{\sf R},\eta}= \sum_{\xi=1}^{16}f_{\eta}^{\xi}\,n_{\xi}.
\end{eqnarray}
Here $m_{{\sf R},\eta}$ measures the occupancy of diamond lattice bonds/pyrochlore lattice 
sites, numbered as in Fig.~\ref{Fig:op_conv}, weighted by the factors $f_{\eta}^\xi$ given 
in Table~\ref{table}. 


Naively, one might have expected the order parameter to have been given by the sum of projectors 
$$\sum_{\zeta=1}^8 \langle {\sf R}_\zeta | \psi \rangle^2$$ 
where $|{\sf R}_\zeta\rangle$ are the 8 possible ordered {\sf R}-states \mbox{($\zeta=1,2,...,8$)}. 
However,
$$\sum_{\zeta =1}^8 \langle {\sf R}_\zeta | \psi \rangle = \frac{N}{4}$$
measures the total number 
of dimers (equal to one quarter of the number of pyrochlore sites), and the only linearly independent combinations of $|{\sf R}_\zeta\rangle $ are 
those given in Table.~\ref{table}.

\section{Evidence for a quantum $U(1)$-liquid phase}
\label{simulation}


\subsection{Comments on simulations}
\label{GFMC}


In this section of the paper, we calculate the ground state properties of the quantum 
dimer model Eq.~(\ref{eq:QDM}) using the Green's Function Monte Carlo 
(GFMC)~\cite{trivedi90, calandra98} method.
This method has previously been applied very successfully to QDM's in two 
dimensions~\cite{ralko05,ralko06,ralko07}.  
We benchmark our GFMC simulations against results from exact diagonalization for small clusters, 
against a perturbation theory about the RK point for \mbox{$\mu \to 1$}  
and, in Appendix~\ref{perturbation}, against expansions about a perfectly ordered 
{\sf R}--state for \mbox{$\mu < 0 $}.  
We chose to use GFMC because it automatically conserves the flux 
quantum numbers which are central to our study of the QDM, 
and because it can provide numerically exact answers 
for the ground state ($T=0$) properties of reasonably large clusters.


GFMC works by calculating a set of running averages on a random
walk through configuration space, using on Hamiltonian matrix elements
to move from one configuration to another.
For large clusters and/or complicated problems, this random walk
must be guided using a trial wave function, previously optimized by a 
suitable variational calculation.  
In this sense GFMC can perhaps best be understood as
a systematic way of improving upon a known variational wave function.


In Appendix~\ref{VMC} we discuss a range of different variational wave functions which
can be used to explore the ground state of the QDM on a diamond lattice.
The most useful guide function we have found to date is~:
\begin{eqnarray}
|\psi^{\sf Var}_{\alpha\beta\gamma}\rangle_{\phi, {\sf c}}  
=\exp [\alpha N_{\sf f} 
+ \beta  m_{\sf R} 
+ \sum_{\langle ij \rangle} \gamma_{ij} \tau_i \tau_j ] 
|\psi_{\text{RK}}\rangle_{\phi, {\sf c}} 
\label{eq:alpha_beta_gamma}
\end{eqnarray}
Here $|\psi_{\text{RK}}\rangle_{\phi, {\sf c}} $ is the equally weighted superposition of all 
dimer configurations within the maximally-connected subsector with flux 
$\vec{\phi} = (\phi_1, \phi_2, \phi_3)$, i.e. those which are connected to an {\sf R}-state 
by string excitations and/or matrix elements of Eq.~(\ref{eq:QDM}).  
The sum $ij$ runs over pairs of hexagonal plaquettes within the diamond lattice, and
$\tau_i = 0$,$1$ is an Ising-like variable which takes on the value $\tau_i=1$ when 
the hexagonal plaquette $i$ is ``flippable''.
$N_{\sf f} = \sum_i \tau_i$ counts the total number of flippable plaquettes and 
$m_{\sf R}$ is the order parameter associated with the {\sf R}--state, as defined in
Eq.~(\ref{eq:op}).
The variational parameters $\alpha$, $\beta$ and $\gamma_{ij}$ are minimized 
using a stochastic reconfiguration algorithm~\cite{sorella01}.  
For large lattices there are a many different pairs of hexagonal plaquettes. 
In practice we restrict ourselves to a set of up to 40 inequivalent $\gamma_{ij}$.


Our GFMC algorithm follows the prescription of~[\onlinecite{calandra98}].  
Simulations are typically performed using up to a few thousands of ``walkers'' in parallel.
The transition probabilities between dimer configurations are calculated 
using the matrix elements from Eq.~(\ref{eq:QDM}).  
If the guide wave function is exact (for example at the RK point, with 
all variational parameters set to zero) GFMC converges to the exact answer
in a single step.  
Away from the RK point simulations typically converge after a few hundreds of 
reconfiguration  steps. 
A ``forward walker'' technique is used to calculate the average value of 
the order parameter $m_{\sf R}$~[\onlinecite{calandra98}].


For $\vec{\phi}=0$, all simulations are started from a perfectly ordered {\sf R}-state, 
and explore only dimer configurations connected with this.
For more general $\vec{\phi}$, starting configurations are constructed using 
the prescription described in Section~\ref{strings}.
In neither case is the GFMC technique ergodic, in the sense of exploring all
dimer configurations which might, in principle, contribute to the ground state.  
However GFMC performed in this way does produce results which are 
representative of the entire configuration space.    
This question is discussed at length in Section~\ref{ergodicity}.


\subsection{Death of an ordered state}
\label{ordered_state}

\begin{figure}[h]
\begin{center}
\includegraphics[width=8cm]{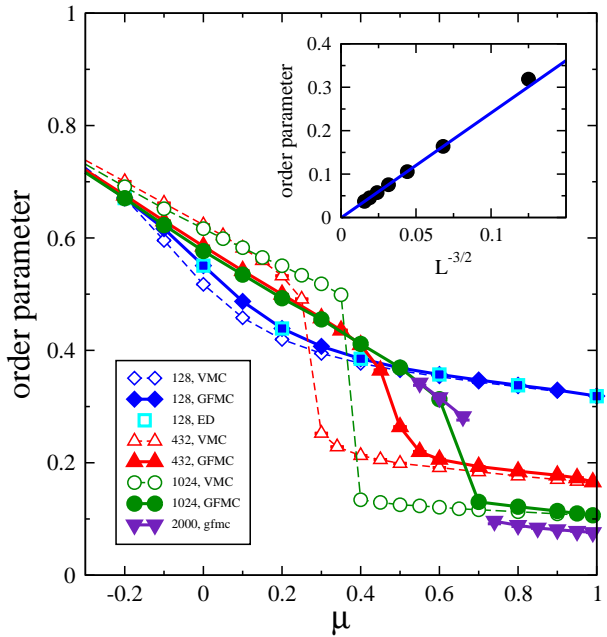}
\end{center}
\caption{
(Color online) 
Collapse of the ordered state: the order parameter $m_{\sf R}$ in 
the sector connected to the {\sf R}-state.  
The results of VMC (dashed lines) and GFMC (solid lines) simulations are shown for 
series of clusters with cubic symmetry, and compared to exact diagonalization 
results for the smallest, 128~bond cluster.  
The inset shows finite size scaling of the order parameter at the RK point.  
 }
\label{Fig:op_cubic}
\end{figure}

We begin by studying how the ordered {\sf R}-state evolves as a function of $\mu$.
In Fig.~\ref{Fig:op_cubic} we present both VMC and GFMC simulation results for the order
parameter $m_{\sf R}$, for \mbox{$-0.2 \leq \mu \leq 1$}, in a series of cubic clusters of the 
form [100] which have the full symmetry of diamond lattice and have $2L^3$ bonds 
\mbox{[cf. Table~\ref{table_g}]}.
The linear dimension of the clusters ranges from $L=4$ (128 bonds)
to $L=10$ (2000 bonds).


\begin{figure}[tbp]
\begin{center}
\includegraphics[width=8cm]{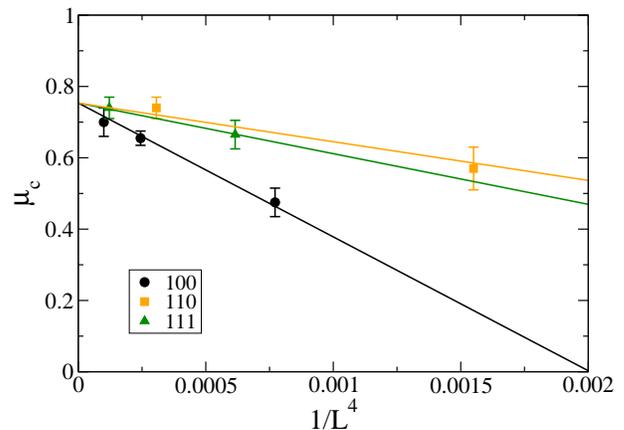}
\end{center}
\caption{
(Color online) 
Estimation of the critical value of $\mu_c(\infty)$ for the transition from the 
ordered {\sf R}-state to the quantum $U(1)$-liquid, taken from GFMC simulation of three
different families of clusters.  
Linear fits to $\mu_c(L)$ as a function of $1/L^4$ are made subject to the constraint that all data 
sets must converge to the same value $\mu_c(\infty)$ for $1/L \to 0$, regardless of the cluster 
geometry.  
We find \mbox{$\mu_c(\infty) = 0.75 \pm 0.02$}. 
}
\label{Fig:fss_muc}
\end{figure}


For the smallest 128-bond cluster we find very close agreement between 
GFMC and VMC simulations, and perfect numerical agreement 
between GFMC results and exact diagonalization calculations.   
This cluster is too small to be representative, but a point of inflection
in $m_{\sf R}$ for $\mu \approx 0$ hints at a possible phase transition
out of the {\sf R}-state.
For larger clusters, a first-order phase transition out of the {\sf R}-state 
is clearly visible as a jump in the order parameter $m_{\sf R}$ at a critical 
value of $\mu = \mu_c(L)$.   
This jump is most pronounced in VMC calculations, where it occurs
for smaller values of $\mu$, i.e. deeper inside the ordered state.
The jump in the $m_{\sf R}$ shifts steadily to larger values of $\mu$
as the size of the system is increased.
For the $L=6$, 432-bond cluster, a jump in $m_{\sf R}$ is observed in VMC
for $\mu_c \approx 0.25$, and in GFMC for $\mu_c \approx 0.45$.
Meanwhile for the largest $L=10$, 2000-bond cluster, the GFMC simulations
suggest that this transition takes place for $\mu_c \approx 0.7$.


For \mbox{$\mu_c(L)  <  \mu < 1$, $m_{\sf R}(\mu)$} takes on a smaller, roughly constant 
value.
However the finite size effects are large, 
with the value of $m_{\sf R}$ decreasing as the system size is increased.
Much larger systems can be simulated at the RK point $\mu \equiv 1$, 
where the form of the ground state wave function is known exactly. 
Various methods of simulating at the RK point are discussed in Section~\ref{ergodicity}.  
In the inset to Fig.~\ref{Fig:op_cubic}, we show results obtained using local updates within 
the configurations explored by GFMC simulations.    
We find that $m_{\sf R}$ vanishes in the thermodynamic limit 
as \mbox{$m_{\sf R}(\mu\equiv 1) \sim L^{-3/2}$}.


Given the strong finite size effects visible in these data, it is important to make an 
estimate of $\mu_c(L \to \infty )$, to ensure that the competing phase for $\mu > \mu_c(L)$
does not collapse to a single point in the thermodynamic limit.  
The properties of the system for $L \to \infty$ {\it must} be independent 
of the cluster geometry, and so $\mu_c(L)$ should interpolate to the same value of 
$\mu_c(\infty)$ for {\it all} different families of clusters.
In Fig.~\ref{Fig:fss_muc} we plot $\mu_c(L)$ as a function of $1/L^4$ for 
clusters of the type [100], [110] and [111], as defined in Table~\ref{table_g}.
All data collapse to a single value $\mu_c(\infty) = 0.75 \pm 0.02$.
We note that the slightly different method of estimation used in 
Ref.~\onlinecite{sikora09} (fitting a separate line to each family of clusters)
gave a very similar result $\mu_c(\infty) = 0.77 \pm 0.02$. 


The fact that \mbox{$\mu_c(\infty) - \mu_c(L)  \sim 1/L^4$} provides strong, albeit
indirect, evidence for the existence of the gapless photon excitations which are 
a signal feature of the quantum $U(1)$-liquid state.  
We can argue as follows :
since the jump in $m_{\sf R}$ implies a first-order, zero-temperature phase 
transition, the finite size correction to $\mu_c$ must follow from the finite size 
corrections to the ground state energy of the two competing phases.
All excitations about the {\sf R}-state are gapped, and finite size corrections 
to its ground state energy should therefore be small.
Assuming that the competing phase is a quantum $U(1)$-liquid with gapless
photon excitations, the finite size corrections to its ground state energy 
will scale as $\sim 1/L^4$, as described by Eq.~(\ref{eq:zero_point}).  
Provided that the ground state energies of both states evolve smoothly near 
$\mu_c(L)$ (which is known to be true from both exact 
diagonalization and quantum Monte Carlo studies), finite size corrections to 
$\mu_c$ should also then scale as $\sim 1/L^4$.


On the strength of the simulation results for $m_{\sf R}$ presented above, it seems very 
reasonable to conclude that the phase diagram proposed in Fig.~\ref{fig:phase_diagram}
is correct, with a single first-order phase transition from an ordered {\sf R}-state to a quantum
$U(1)$-liquid occurring for $\mu_c = 0.75 \pm 0.02$.
In what follows, we present a variety of other evidence that this is indeed the case.


\subsection{Collapse of monopole string tension}


\begin{figure}[tbp]
\begin{center}
\includegraphics[width=8cm]{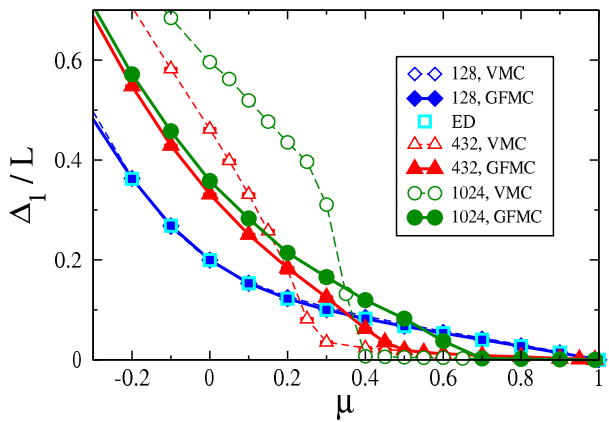}
\end{center}
\caption{
(Color online) 
Monopole string tension $\kappa_1$ [Eq.~\ref{eq:string_tension}] 
for a range of $\mu$ spanning the ordered {\sf R}-state and proposed 
quantum $U(1)$-liquid phase.  
Results are taken from VMC (dashed lines) and GFMC (solid lines) simulations 
for a series of clusters with cubic symmetry, and compared to exact diagonalization 
results for the smallest, 128~bond cluster.
}
 \label{Fig:gap_cubic}
\end{figure}


If the state occurring for \mbox{$\mu_c(L)  <  \mu < 1$} is indeed a quantum $U(1)$-liquid, 
it should possess gapped, deconfined monopole excitations~\cite{moessner03}.
We can study the motion of these monopoles indirectly through the thought experiment used 
to construct configurations in different flux sectors in Section~\ref{strings}.   
As illustrated in Fig.~\ref{fig:string}, separating the two monomers (monopoles) 
created by removing a dimer, destroys flippable hexagons along the ``string" of dimers
connecting the monomers.   
In the crystalline state, where the potential energy dominates, one can expect that there 
is a finite energy cost of such an operation, proportional to the length of the string.   
Where the string defect threads a periodic boundary of the cluster, its energy cost
can be calculated as the gap to the ground state with one additional unit of flux.  


We therefore define the ``string tension" as the energy difference 
between the ground state in the zero-flux sector ($E_0$) and a sector with a single 
string defect ($E_1$), divided by the length of the string ($L$)
\begin{eqnarray}
\kappa_1 = \frac{E_1 - E_0}{L}  =\frac{\Delta_1}{L} 
\label{eq:string_tension}
\end{eqnarray}
For the string defects considered, $L$ is simply the linear 
dimension of the cluster.  
This string tension is an indirect measure of the attractive, confining 
force between monopole excitations.  
It must therefore vanish in the deconfined quantum $U(1)$-liquid, 
but will be finite in the {\it any} ordered state. 


In Fig.~\ref{Fig:gap_cubic} we plot simulation result for the monopole string tension
$\kappa_1$ as a function of $\mu$ for the same set of clusters and 
range of parameters as Fig.~\ref{Fig:op_cubic}.  
Deep inside the ordered state, for negative values of $\mu$, 
classical effects dominate.
A single string defect in a perfectly ordered {\sf R}-state destroys $2L$ flippable 
hexagons at a potential energy cost of $2L|\mu|$.
This linear relationship between $\kappa_1$ and $|\mu|$ for $\mu \lesssim 0$ 
is clearly visible in GFMC simulation results for the larger clusters.
However the string tension drops rapidly as we approach the transition from 
the {\sf R}-state into the competing phase at $\mu = \mu_c(L)$, 
and for large clusters it is essentially zero for $\mu_c(L)  <  \mu \leq 1$.  


The coincidence between the jump in the order parameter $m_{\sf R}$ 
[Fig.~\ref{Fig:op_cubic}] and the collapse in the string tension $\kappa_1$ 
[Fig.~\ref{Fig:gap_cubic}] lends yet more support to the proposed
form of the phase diagram for the QDM on a diamond lattice [Fig.~\ref{fig:phase_diagram}].  
In particular, the absence of a string tension for \mbox{$\mu_c(L)  <  \mu < 1$}
rules out the possibility that the competing phase is a different ordered state, 
for example a three-dimensional analogue of the resonating plaquette phase 
found in quantum dimer and quantum ice models in two 
dimensions~\cite{rokhsar88, ralko08, shannon04}.
However we still need to rule out the possibility that the competing phase is a 
different form of quantum liquid, by confirming explicitly that it is indeed the quantum 
$U(1)$-liquid we have been seeking.  
This is accomplished below.


\subsection{Birth of a $U(1)$-liquid}
\label{liquid}


\begin{figure}[tbp]
\begin{center}
\includegraphics[width=8cm]{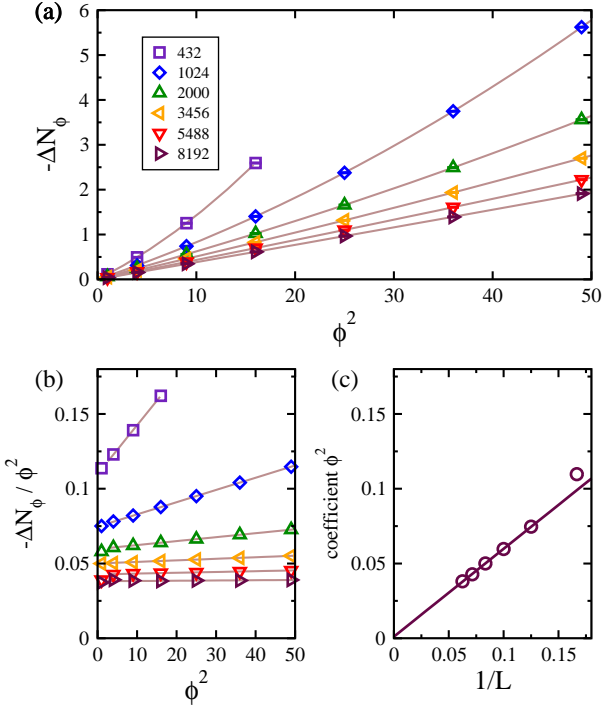}
\end{center}
\caption{
(Color online) 
Reduction in the average number of flippable plaquettes $\Delta N_\phi$, as 
calculated in simulations at the RK point.   
(a) $-\Delta N_\phi$ plotted as a function of $\phi^2$ for cubic clusters of 
linear dimension $L=4$ (432 bonds) to $L =16$ (8192 bonds).   
(b) The same data set plotted as $-\Delta N_\phi/\phi^2$ vs $\phi^2$ to 
extract the leading dependence on~$\phi$.  
(c) Finite-size scaling of the the coefficient of $\phi^2$ found from fits
to $-\Delta N_\phi$ for clusters of linear dimension $L$ 
[cf. intercept to ordinate axis in (b), above].
The linear collapse of the coefficient of $\phi^2$ with $1/L$
is consistent with the prediction of the effective field theory 
for a quantum $U(1)$ liquid Eq.~(\ref{eq:gaps}).  }
\label{fig:DeltaNf}
\end{figure}


\begin{figure}[tbp]
\begin{center}
\includegraphics[width=8cm]{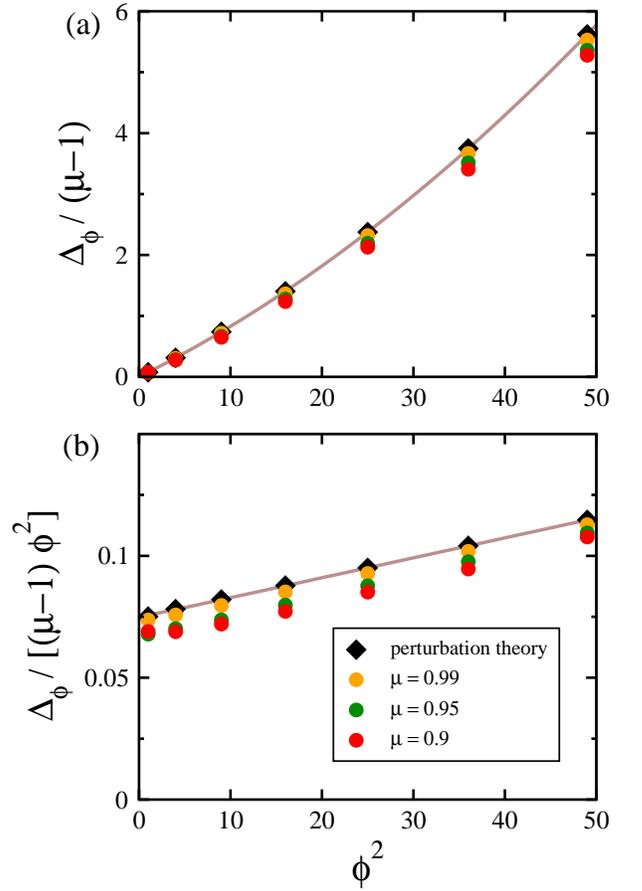}
\end{center}
\caption{(Color online) 
Comparison of the energy gaps $\Delta_{\phi}$ found in GFMC simulations 
with the predictions of a perturbation theory about the RK point [Eq.~(\ref{eq:pert})],
for a range of $\mu$ within the proposed quantum $U(1)$-liquid phase.  
(a) Energy gap $\Delta_{\phi}$ normalized to \mbox{$-\delta \mu = \mu-1$}, 
to extract the leading dependence on $\mu$, and plotted as 
a function of $\phi^2$ [cf. Eq.~(\ref{eq:gaps})].  
(b) Energy gap $\Delta_{\phi}$ normalized to \mbox{$-\delta \mu \times \phi^2$}, 
to extract the leading dependence on both $\mu$ and $\phi$, plotted as a 
function of $\phi^2$ [cf. Eq.~(\ref{eq:gaps})].  
All results are for a 1024-bond cluster with cubic symmetry.
 }
\label{fig:near_rk}
\end{figure}


Within the field-theoretical scenario for a three dimensional QDM on a bipartite 
lattice, the quantum $U(1)$-liquid phase ``grows'' out of the RK 
point~\cite{moessner03,bergman06-PRB}.   
Exactly at the RK point, the ground states in all flux sectors are degenerate.   
Away from the RK point, in a finite size system, the ground state in the zero-flux 
sector has the lowest energy, and the (finite-size) energy gaps to the other low-lying
flux sectors scale as $\Delta_\phi \sim c^2 \phi^2/L$ [cf. Eq.~(\ref{eq:gaps})].
By construction, $c$ vanishes at the RK point.   


Since the exact ground state is known at the RK point, we can calculate the 
finite size gaps $\Delta_\phi$ (and implicitly the speed of light $c$) within
perturbation theory about the RK point.   
Writing 
\begin{eqnarray}
{\mathcal H}_{\sf QDM} = {\mathcal H}_{\sf RK} - \delta \mu \,\mathcal{N}_{\sf f}
\label{eq:pert}
\end{eqnarray}
where ${\mathcal H}_{\sf RK}$ is defined by Eq.~(\ref{eq:RK}), 
$\delta \mu = 1-\mu$,  and the operator $\mathcal{N}_{\sf f}$ is defined 
by Eq.~(\ref{eq:Nf}), we find 
\begin{eqnarray}
\Delta_{\phi, {\sf MC}} =  \Delta N_{\phi, {\sf MC}}   \delta\mu + {\mathcal O}(\delta\mu^2)
\label{eq:deltaE}
\end{eqnarray}
where 
\begin{eqnarray}
\Delta N_{\phi, {\sf MC}}   
=  \langle {\mathcal N}_{\sf f} \rangle^{\sf RK}_{\phi, {\sf MC}} 
 - \langle {\mathcal N}_{\sf f} \rangle^{\sf RK}_{0, {\sf MC}} 
\label{eq:deltaN}
\end{eqnarray}
is the difference in the average number of flippable plaquettes between the zero-flux sector 
and the sector with flux $\phi$, calculated at the RK point, within the maximally-connected
subsector of the Hilbert space, $\lambda_c= {\sf MC}$.   
The maximally flippable {\sf R}-state belongs to the zero-flux sector, and
we therefore anticipate that $\Delta N_{\phi} < 0$.   
If, as supposed, a quantum $U(1)$-liquid grows adiabatically from the RK
point, then we must find 
\begin{eqnarray}
\Delta N_{\phi}  \sim - \frac{\phi^2}{L}
\label{eq:Nphi}
\end{eqnarray}
%


In Fig.~\ref{fig:DeltaNf} we show results for $\Delta N_{\phi}$ obtained from simulation 
of a series of [100] cubic clusters with linear dimension ranging from $L=6$ (432 bonds) to 
$L=16$ (8192 bonds), for flux $\vec\phi = (\phi, 0, 0)$,  plotted as a function of $\phi^2$.   
Simulations were performed within the maximally-connected subsector of states connected 
to the {\sf R}-state for $\phi=0$, and within the subsector associated with a given number of 
string defects for higher values of flux.  
Due to large finite size effects, the expected linear relationship between $-\Delta N_{\phi}$ and $\phi^2$ is only visible for clusters with more than 2000 bonds.  
However the coefficient of $\phi^2$ can be extracted from a polynomial fit to 
$\Delta N_{\phi}$ as a function of $\phi$, at given $L$.  
This is found to be proportional to $1/L$, as shown in the inset to Fig.~\ref{fig:DeltaNf}.
From these results we obtain 
\begin{eqnarray}
c^2\approx 0.6 \times \delta \mu
\end{eqnarray}
(in the present units) confirming that an incipient quantum $U(1)$-liquid is present 
at the RK point.   
This result is specific to the QDM on a diamond lattice.  
We note that simulations of the QDM on a cubic lattice yield quite different 
results~\cite{olga-unpub}.


In Fig.~\ref{fig:near_rk} we compare the results of the perturbation theory Eq.~(\ref{eq:deltaE}) 
with GFMC calculations made for a set of parameters bordering on the RK point, 
for a cluster of 1024 diamond lattice bonds.
The energy gaps obtained from GFMC are plotted as $\Delta_{\phi}/(1-\mu)$ 
so to compare with perturbation theory, and as a function of $\phi^2$ 
so as to extract the leading behaviour of the $U(1)$-liquid phase.  
Both sets of data are found to show the same $\phi^2$ dependence on flux sector, 
confirming the existence of a quantum $U(1)$-liquid bordering the RK point.   


\subsection{A phase transition in a spectrum}
\label{spectrum}


\begin{figure}[tbp]
\begin{center}
\includegraphics[width=8cm]{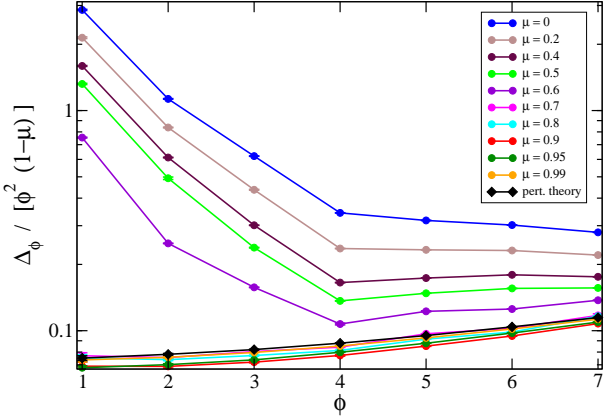}
\end{center}
\caption{
(Color online) 
First order quantum phase transition from the ordered {\sf R}-state into the quantum 
$U(1)$-liquid, as imaged in the finite size spectrum of a 1024-bond cluster.   
GFMC simulation results for the energy gap $\Delta_{\phi}$ are plotted as 
$\Delta_{\phi}/[\phi^2(1-\mu)]$ so as to collapse the spectra for the $U(1)$-liquid phase.  
The phase transition is visible as an abrupt qualitative change in the form of the 
spectra for $\mu = 0.7$.  
The results of perturbation theory about the RK point (black line and points) 
are shown for comparison.
}
\label{Fig:f00_gfmc}
\end{figure}


We are now in a position to draw a set of strong conclusions about the zero-temperature 
phase diagram of the QDM on a diamond lattice 
(cf. Fig.~\ref{fig:phase_diagram}).  
For $$\mu < \mu_c = 0.75 \pm 0.02$$ the ground state of the model is an ordered {\sf R}-state.
For \mbox{$\mu_c  < \mu < 1$} it is a quantum $U(1)$-liquid with linearly dispersing
photon excitations and deconfined magnetic monopoles.   
The quantum phase transition between these two phases at $\mu = \mu_c$ is first order.
For $\mu \to 1_-$, the quantum $U(1)$-liquid is adiabatically connected with the RK point.


In Fig.~\ref{Fig:f00_gfmc} we present the results of GFMC simulations of a cubic 
1024-bond cluster, for a range of parameters \mbox{$0 \leq \mu \leq 1$} spanning 
both the {\sf R}-state and the quantum $U(1)$-liquid.  
We plot the energy gaps $\Delta_\phi$, rescaled as $\Delta_{\phi}/[\phi^2(1-\mu)]$ 
so as to collapse all data within the $U(1)$-liquid phase.  
Deep within the ordered phase the $\Delta_\phi$ should be proportional 
to both the linear dimension of the cluster and the flux (number of string defects), 
i.e. $$\Delta_ \phi^{\sf R} \sim L\times\phi$$


A plot of $\Delta_ \phi/\phi^2$ at fixed $L$ should therefore show a clear distinction 
between a $U(1)$-liquid ($\Delta_ \phi/\phi^2 \sim \text{const.}$), and an ordered 
phase ($\Delta_ \phi/\phi^2 \sim 1/\phi$).   
For this cluster size a clear division is observed between $\mu \ge 0.7$ (liquid) 
and $\mu \le 0.6$ (ordered).
This is unambiguous evidence of a phase transition from a linearly confining 
phase to a quantum $U(1)$-liquid, imaged in a spectrum.  
The abrupt, qualitative change in the spectra for $\mu = 0.6$ confirms that this 
phase transition is first-order in character.  


\section{Ergodicity and conservation laws within a given flux sector}
\label{ergodicity}


\subsection{Lessons from exact diagonalization}


Since GFMC uses on matrix elements of the QDM Hamiltonian [Eq.~(\ref{eq:QDM})] to 
connect different dimer configurations, it is guaranteed to respect the flux quantum numbers 
defined in Section~\ref{strings}.
However, not all dimer configurations with the same magnetic flux are connected by 
matrix elements of the Hamiltonian.
Clearly, therefore, GFMC simulations for a given flux are {\it not} ergodic, in the sense 
that they do not explore all dimer configurations which might, in principle, contribute to the 
ground state for that flux sector.
It is therefore very important to consider how the Hilbert space for a given flux $\vec\phi$
breaks up into different subsectors, and to ensure that simulations are carried out in the 
appropriate sub-sector(s).
To this end, much can be learnt from the exact diagonalization of small clusters.


\begin{figure}[tbp]
\begin{center}
\includegraphics[width=8cm]{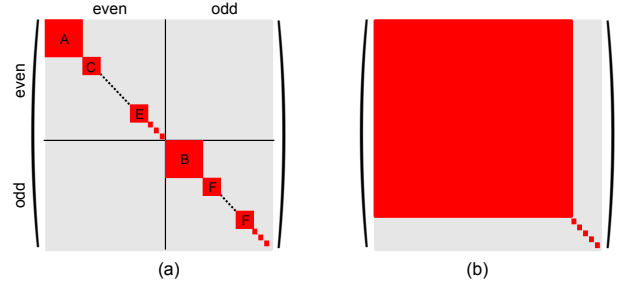}
\end{center}
\caption{
(Color online) 
(a) Block-structure of the Hamiltonian for the quantum dimer model on 
a diamond lattice, for an 128-bond cluster with a fixed value of the flux 
$\vec{\phi} = (0,0,0)$, following \mbox{Table~\ref{sectors}}.  
Connected dimer configurations break up into different sub-ensembles.
These can be divided into two groups, according to whether their transition
graphs contain an odd or even number of loops.
Both even and odd subsectors contain isolated configurations which do not
participate in the dynamics of the model, and do not contribute to the ground state 
for $\mu < 1$.   
(b) Contrasting block-structure of the Hamiltonian for the quantum dimer model 
on a square lattice for a finite size cluster with a fixed value of the flux 
$\vec{\phi} = (0,0)$.  
All dimer configurations which contribute to the ground state for $\mu < 1$ 
belong to a single connected block.
The remaining, isolated, dimer configurations do not participate in the 
dynamics of the model.
}
\label{Fig:mondrain}
\end{figure}


In total there are 115,150,848 dimer coverings of the $L=2$, 128-bond, cubic cluster.
Of these, 11,835,352 have zero flux ($\phi=0$).
These zero-flux configurations include 1440 isolated configurations with no
flippable plaquettes, and 3072 configurations with a single flippable plaquette 
which connect only to a single other configuration.
The remaining 11,830,840 configurations can be grouped into a small number of 
subsectors which are connected by the matrix elements of Eq.~(\ref{eq:QDM}).
These are listed in Table~\ref{sectors}, and illustrated schematically in 
Fig.~\ref{Fig:mondrain}(a).  
This situation should be contrasted with the much simpler structure of the Hilbert
space in the QDM on a square lattice, where all connected dimer configurations 
for a given value of flux $\vec\phi=(\phi_x,\phi_y)$ are associated with a single 
block of the Hamiltonian --- cf. Fig.~\ref{Fig:mondrain}(b).  


\begin{table}[h!]
\caption{
Subsectors of the Hilbert space of the QDM on a diamond lattice
with zero flux ($\phi=0$) for a cubic, 128-bond cluster.  
Subsectors are listed in order of increasing dimension, with isolated configurations 
and subsectors consisting of only two configurations omitted.   
The largest, maximally-connected, subsector (A) contains the \mbox{{\sf R}-states}. 
Further subsectors can be obtained through cyclic permutations of dimers on 
closed loops --- the \mbox{8-link} loop operation shown in Fig.~\ref{fig:loop8} 
provides access to subsector B.
Subsectors can be classified according to the number and type of loops 
in their transition graph.
} \begin{center}
\begin{tabular}{|c|c|c|c|c|}
\hline\hline
\begin{tabular}{c}
subsector\\
label\\
$\lambda_c$\\
\end{tabular}
& 
multiplicity 
& 
dimension  
&  
\begin{tabular}{c}
number of \\
loops in \\
transition \\
graph (mod 2) 
\end{tabular} 
& 
\begin{tabular}{c}
longest\\
irreducible \\
loop in \\
transition graph
\end{tabular} 
\\
\hline\hline
A &1 & 7794744 &0 & 2 \\
B &1 & 3468032 &1 & 8\\
C & 3 & 131584 & 0& 10\\
D & 3 & 45184 & 0& 14 \\
E & 8 & 3376 & 0 & 14\\
F &192 & 40 &1& 12\\
\hline\hline
\end{tabular}
\end{center}
\label{sectors}
\end{table}%
  
\begin{figure}[tbp]
\begin{center}
\includegraphics[width=8cm]{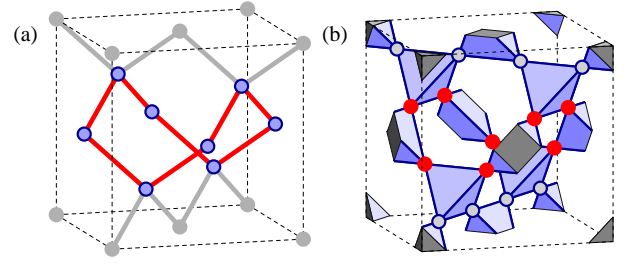}
\end{center}
\caption{
(Color online) 
8-link ``loop'' used to connect dimer configurations within subsectors A and B 
of the Hilbert space of the QDM on a diamond lattice [cf. Table~\ref{sectors}]. 
(a) 8-link loop of bonds within the cubic unit cell of the 
diamond lattice.
Three distinct 8-link loops of this type are possible, each of which 
appears octagonal when projected into an appropriate [100] plane.
Where exactly 4 bonds are occupied by dimers, 
this loop is ``flippable'', i.e. a new state obeying 
the dimer constraint can be obtained by cyclicly 
permuting those dimers around the loop.
(b) equivalent 8-link loop of sites within the cubic unit 
cell of the pyrochlore lattice.
}
\label{fig:loop8}
\end{figure}

For this 128-bond cluster, about two-thirds of the configurations with $\phi=0$ 
belong to a maximally-connected subsector, labelled A in Table~\ref{sectors}, 
which includes the \mbox{{\sf R}-states}. 
The vast majority of the remaining configurations --- about one-third of the total --- are
found in one other subsector, labelled B in Table~\ref{sectors}.
Subsector B can be accessed from the maximally connected subsector by acting on a 
randomly chosen configuration with the update shown in Fig.~\ref{fig:loop8} --- the cyclic 
permutation of dimers around a closed, 8-link loop.
Interestingly, this 8-link permutation is the sub-leading term for dimer dynamics in the 
derivation of the QDM from Eq.~(\ref{eq:tVmodel}) or Eq.~(\ref{eq:XXZmodel}).
%


Subsectors C, D, E and F {\it cannot} be accessed from subsector A using the 8-link 
loop alone.  
However they {\it can} be reached by cyclically permuting dimers within suitably chosen 
configurations belonging to subsector A, on loops of length 10, 14, 14 and 12, respectively.  
Since the number and length of possible such loops grows with the size 
of the system, we anticipate that the number of subsectors 
in the Hilbert space grows with system size. 
We put these observations on a more formal footing below.


\subsection{A hidden quantum number}


The way in which the Hilbert space of the QDM on a diamond lattice 
breaks up into distinct blocks suggests the existence of further, conserved quantities 
besides magnetic flux $\phi$.
We can gain more insight into this problem, and understand the action of the 8-link loop
shown in Fig.~(\ref{fig:loop8}) by studying transition graphs.  


\begin{figure}[tbp]
\begin{center}
\includegraphics[width=5cm]{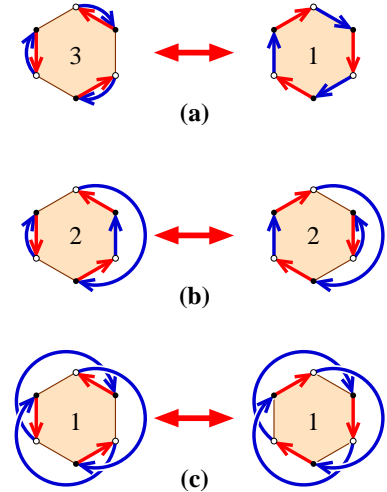}
\end{center}
\caption{
(Color online) 
Schematic representation of all possible changes in the topology of a transition graph
which follow from the cyclic permutation of dimers on a hexagonal 6-link loop.
Red lines denote loop segment with a single dimer.
Blue lines denote a loop segment with an odd number dimers. 
All loop segments, whether red or blue, start and finish on sites belonging to 
different sublatticies of the (bipartite) diamond lattice, denoted here by filled 
or empty circles.
The number of loops associated with each section of the transition graph is shown 
in the middle of the hexagonal plaquette.
This number changes by either 0 or 2.
}
 \label{fig:transition6}
\end{figure}


\begin{figure}[tbp]
\begin{center}
\includegraphics[width=5cm]{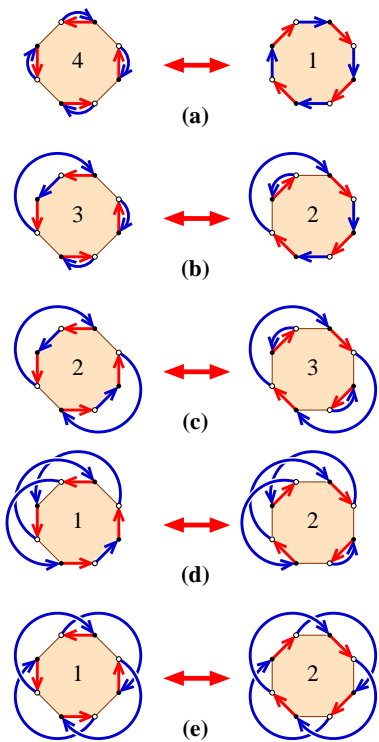}
\end{center}
\caption{
(Color online) 
Schematic representation of all possible changes in the topology of a transition graph
which follow from the cyclic permutation of dimers on an 8-link loop.
The number of loops of the transition graphs changes by either 1 or 3.
}
 \label{fig:transition8}
\end{figure}


A transition graph is the set of directed loops formed by overlaying two dimer configurations 
on a bipartite lattice, and drawing arrows from A to B sublattice on those bonds which are
occupied by a dimer in the first configuration, and from B to A sublattice on those bonds which 
are occupied by dimers in the second configuration~\cite{rokhsar88}.
Cyclicly permuting the dimers on the loops of the transition graph converts the first configuration 
into the second (and vice versa).
By construction, these loops cannot intersect one another --- although one loop may be threaded 
through another in a three-dimensional lattice.


A pair of dimer configurations are connected if (and only if) we can undo all the long loops in their 
transition graph using only matrix elements of Eq.~(\ref{eq:QDM}) --- cyclic permutations of 
dimers on a 6-link hexagonal plaquette --- so that only trivial ``loops'' occupying 
a single bond remain.  
These trivial loops have length 2, and the maximum number of trivial loops achievable 
is equal to the number of dimers in the system --- an even number for all of the clusters 
considered in this paper.  
In Fig.~\ref{fig:transition6} we enumerate the three possible, topologically distinct, changes in
the transition graph associated with the cyclic permutation of dimers on an hexagonal plaquette. 
In each case the total number of loops in the transition graph changes by an even number
--- either 0 or 2.


Cyclic permutations of dimers on an 8-link loop 
have a somewhat richer structure, illustrated in Fig.~\ref{fig:transition8}.   
In this case, the total number of loops in the transition graph changes by an odd number --- 
either 1 or 3.
Consequently, the action of 8-link loop cannot be undone using matrix elements of 
the Hamiltonian [Eq.~(\ref{eq:QDM})] alone.


More formally, we can associate a unique transition graph with each distinct dimer 
configuration by choosing one particular dimer configuration as reference
(for present purposes, an \mbox{{\sf R}-state} is a convenient choice).
The number of loops in this transition graph, modulo 2, is then a conserved 
quantity within the dynamics of the QDM Eq.~(\ref{eq:QDM}).
Furthermore, it must be even in the subsector connected with the \mbox{{\sf R}-state}.
In Table~\ref{sectors} we classify the all of the different subsectors of the zero-flux
sector of the 128-bond cubic cluster containing more than two configurations 
according to this new, Ising, quantum number.


Examination of transition graphs also provides some insight into the other, smaller, 
subsectors of the Hilbert space, labelled C, D, E and F in Table~\ref{sectors}.  
Empirically, we find that these can be accessed from the maximally-connected subsector 
containing the \mbox{{\sf R}-states} (subsector A) through the cyclic permutation of dimers 
on suitably chosen loops of length 10, 12, and 14.
Closed loops in the transition graph can generally be contracted through the repeated
permutation of dimers on 6-link loops of the type found in the Hamiltonian Eq.~(\ref{eq:QDM}).
By construction, {\it all} non-trivial loops in the transition graph for two configurations within 
subsector A can be removed in this way, leaving only trivial loops of length 2.
The different subsectors B---F listed in Table~\ref{sectors} can therefore be classified 
according to the longest irreducible loop in their transition graph with a (suitably chosen)
state within subsector~A.
The lengths of the irreducible loops are listed in Table~\ref{sectors}.


It is interesting to note that the Ising quantum number associated with the number
of loops in the transition graph is even where the longest irreducible loop is of
length 2, 10 or 14, and odd where the longest irreducible loop is of length 8 or 12.
We infer (but have not proved) that topological selection rules of the form illustrated in 
Fig.~\ref{fig:transition6} apply for updates based on loops of length 6, 10, 14\dots, and 
of the form illustrated in Fig.~\ref{fig:transition8} for updates based 
on loops of length 8, 12, 16\dots  
Cyclic permutations on loops of length 6, 10, 14\ldots 
are precisely the form of dynamics permitted for (spinless) fermionic 
models of the form Eq.~(\ref{eq:tVmodel}), while 
bosonic models also permit cyclic permutations on loops of length 8, 12, 16\ldots .\cite{schwandt10}
We therefore speculate that the number of loops in the transition graph will be a conserved
quantity (mod 2) for the most general QDM Hamiltonian derived from a fermionic model
on a pyrochlore lattice, but {\it not} for the equivalent bosonic or spin model.


Clearly, more questions could be asked about the non-ergodicty of the QDM on a diamond lattice.
What, for example, are the the conserved quantities which distinguish 
the smaller subsectors of the Hilbert space ?
And how do the number, multiplicity, and dimension of these subsectors grow with the size
of the system ?
For the purposes of this paper, however, the most pressing need is to understand the impact of 
this non-ergodicity on the simulations described in Section~\ref{simulation}.
It is this to which we now turn.  
  

\subsection{Can we trust simulations ?}


It is clear from the analysis above that QDM on a diamond lattice is {\it not} ergodic
within a given flux sector.    
The analysis which leads to the phase diagram Fig~\ref{fig:phase_diagram} rests on 
simulations performed within the (maximally connected) subsector of states connected 
to the {\sf R}-state by the matrix elements of Eq.~(\ref{eq:QDM}), and on the systematic 
construction of configurations with finite flux using ``string'' defects' [cf. Fig.~(\ref{fig:string})].
Before we can have full confidence in these results
we must therefore address the following question~:


{\it What bias do we introduce into our results by simulating only within these subsectors
of the Hilbert space~?}


The enumeration of states for an 128-bond cluster confirms the common-sense 
expectation that we can access the largest subsector of the Hilbert
space for zero flux by seeding simulations from an \mbox{{\sf R}-state}.
Moreover, we now understand how to extend these simulations to configurations 
with an odd number of loops in their transition graph, through use of an 8-link loop.
However there remain a non-vanishing fraction of dimer configurations  --- about 5\% for 
the 128-bond cubic cluster [Table~\ref{sectors}] --- which can only be reached by other, 
more complex updates.


It seems reasonable to expect that these unsociable configurations will (a) contribute 
little to the ground state wave function and (b) constitute a vanishing fraction
of the total number of dimer configurations in the thermodynamic limit.
However this reasonable expectation must be approached with some caution, 
not least because, in the absence of a full understanding of conserved quantities, 
it is hard to assess how these neglected corners of the Hilbert space grow with 
system size.


We can most easily get a handle on the consequences of non-ergodicity at the RK point, 
where simulations for ground state properties amount to performing 
classical averages over all dimer configurations within a given flux sector.   
These can be performed in several different ways
\begin{enumerate}
\item As in Section~\ref{liquid}, using ``local'' updates based on the matrix 
         elements of the Hamiltonian.  
\item Using a ``loop'' algorithm developed for the simulation of ice 
         models~\cite{newman07,hermele04}.   
\item Using a ``worm'' algorithm based on movement of monomers~\cite{barkema98}.  
\end{enumerate}


In the ``loop'' algorithm, a directed random walk along bonds 
is used to construct a self-intersecting path in which all flux arrows
associated with dimers have the same sense.  
The dimers which lie along this path are each cycled one link
further along, reversing the sense of the flux around the closed ``loop'', 
while preserving the dimer constraint.   


In the ``worm'' algorithm, a single dimer is removed, and the two 
monopoles created are allowed to diffuse freely around the lattice
until they meet, at which point the missing dimer is re-introduced.  
We have checked explicitly that ``loop'' and ``worm'' simulations lead to 
identical results for the QDM on the diamond lattice, and concentrate here  
on comparing results obtained using ``local'' and ``loop'' updates.  

\begin{figure}[tbp]
\begin{center}
\includegraphics[width=8cm]{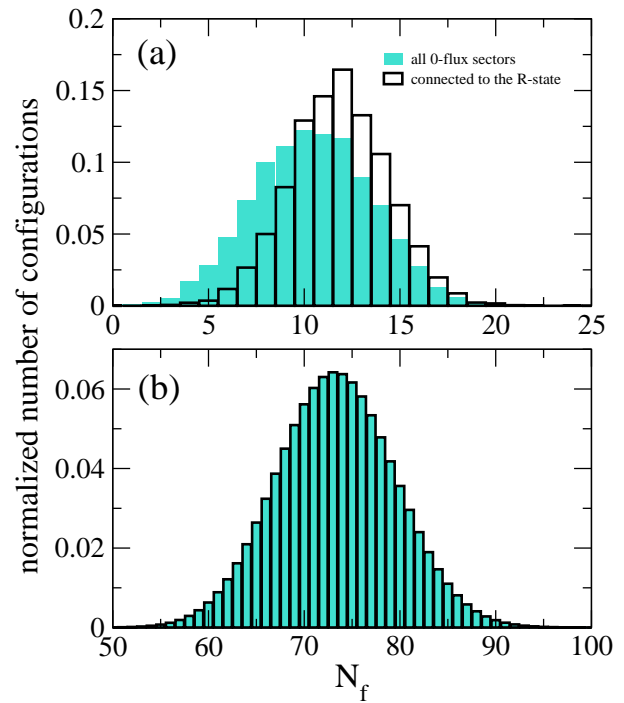}
\end{center}
\caption{
(Color online) 
Histogram of the number of flippable plaquettes in different sets of dimer configurations, 
calculated at RK point.
(a) Results for an 128-bond cubic cluster, calculated within in the entire zero-flux sector 
(blue, shaded columns), and in the subsector connected to the \mbox{{\sf R}-state} 
(unshaded columns).
(b) Equivalent results a for an 1024-bond cluster.
}
 \label{fig:histograms_Nf}
\end{figure}

A necessary condition for the existence of a quantum $U(1)$-liquid in a QDM 
is that the average number of flippable plaquettes in the dimer configurations 
belonging to a given flux sector scale as 
$\Delta N_{\phi} = N_{\sf f}(\vec\phi) - N_{\sf f}(\vec{0}) \sim - \phi^2/L$ 
[Eq.~(\ref{eq:Nphi})].   
In Fig.~\ref{fig:histograms_Nf}(a) we compare the probability
distribution for the number of flippable plaquettes $N_{\sf f}$ calculated
for all dimer configurations within the zero-flux sector (green, shaded columns) 
with that calculated within the maximally 
connected subsector containing the {\sf R}-state (columns without shading).
Identical results were obtained both within simulation, using ``loop'' updates to explore 
all dimer configurations, and ``local'' updates to access to the maximally-connected 
subsector, and within exact diagonalization, by explicit enumeration of all dimer 
configurations. 
The probability distribution for the number of flippable plaquettes is markedly
different in the two cases, and the mean value of $N_{\sf f}$ clearly higher 
for the subset of configurations connected with the {\sf R}-state.
%


At first sight this disagreement might seem rather alarming, but it is simply a finite-size effect.   
In Fig.~\ref{fig:histograms_Nf}(b) we present the results of 
an identical study for a larger, 1024-bond cluster.
The same, Gaussian, probability distribution with mean 
$\langle N_{\sf f} \rangle= 73.27$ and variance $\sigma_{N_{\sf f}} = 6.21$
is found for both the full set of dimer configurations sampled by 
the ``loop'' algorithm, and the maximally-connected subsector explored 
by the ``local'' updates.
Similar gaussian probability distributions are found in sectors of the Hilbert 
space with finite flux, and are again independent of the method of simulation.


There are two obvious possible explanations as to why the same probability 
distribution should be found for averages calculated at the RK point 
in the full set of dimer configurations, and those calculated within its 
maximally-connected subsector.
One is that the relative size of the maximally-connected subsector grows with system 
size, so that it comes to dominate all averages in the thermodynamic limit.  
The other is that all large blocks of the Hilbert space (within a given flux sector)
are statistically very similar.   
This might occur if, for example, configurations in different subsectors differed only by 
some local defect which could not be removed using matrix elements of the Hamiltonian, 
but which had no impact on the bulk properties of the state.  


\begin{figure}[tbp]
\begin{center}
\includegraphics[width=8cm]{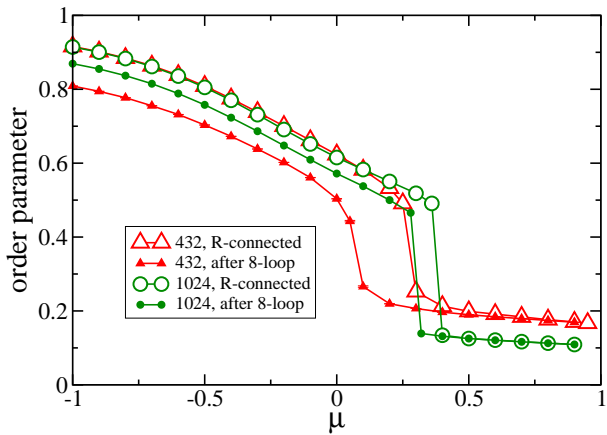}
\end{center}
\caption{
(Color online) 
$\mu$-dependence of the order parameter $m_{\sf R}$ for the {\sf R}-state, within two 
different subsectors of the Hilbert space: starting from the {\sf R}-state in flux $\phi=0$ 
subsector (open symbols) and in the subsector that is connected to it by an 8--site loop 
(solid symbols).   
Results are calculated using Variational Monte Carlo simulation for cubic clusters 
with 432 (triangles) and 1024 (circles) diamond-lattice bonds.
}
\label{fig:VMC_study_ergodicity}
\end{figure}


To distinguish between these alternatives we can examine how quantum Monte Carlo 
simulation results in the two largest subsectors of the Hilbert space 
depend on the subsector ${\lambda_c}$  in which they are carried out.
In Fig.~\ref{fig:VMC_study_ergodicity} we present the results of a Variational Monte Carlo 
study of the order parameter $\langle m_{\sf R} \rangle_{\lambda_c}$,  
for a cubic 432 and 1024-bond clusters.
Simulations were carried out within
\begin{enumerate}[i)]
\item the subsector with flux $\phi=0$ containing the {\sf R}-state;
\item the subsector with flux $\phi=0$ obtained by acting with an 
8-link loop update on a randomly chosen state from the subsector containing the {\sf R}-state.
%
%
\end{enumerate}
%


Within the crystalline, ordered phase, an 8--loop update reduces the 
$m_{\sf R}$ relative to the subsector containing the {\sf R}-state.
This reduction in order parameter scales as $1/L^3$ --- as would be expected
for a local defect into an ordered state --- and vanishes in the thermodynamic 
limit.
Meanwhile the energy cost of the 8--loop update is a constant in the ordered state.
This behaviour should be compared with the topological string defect used to change 
flux sector acting in the  {\sf R}-state ordered phase, whose effect on the order parameter scales as $1/L^2$, and which 
creates a finite size energy gap $\propto L$.


In marked contrast, the simulation results for 
$\langle m_{\sf R}(\mu) \rangle$, in the quantum $U(1)$-liquid phase
are independent of the subsector within which they are calculated.
Intuitively, we imagine that as the {\sf R}-state order melts into the $U(1)$-liquid 
phase, so defects in the order parameter melt, too, leaving only 
an invisible partition in the Hilbert space.


\begin{figure}[tbp]
\begin{center}
\includegraphics[width=8cm]{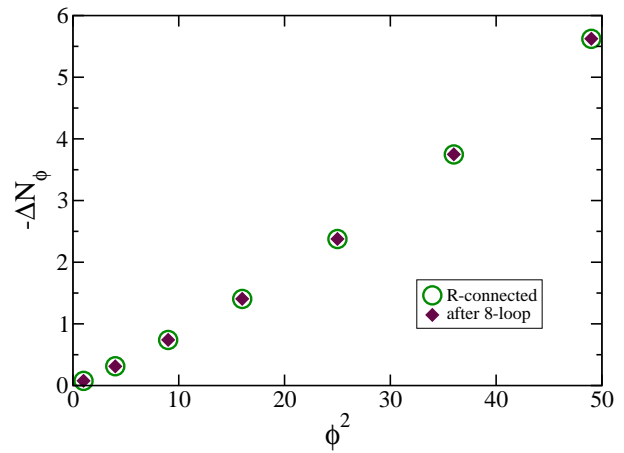}
\end{center}
\caption{
(Color online) 
The difference in the average number of flippable plaquettes $\Delta N_\phi$
[Eq.~(\ref{eq:deltaN})],  calculated at the RK point for dimer configurations in 
different subsectors of the Hilbert space.  
Green circles denote results for configurations connected to the 
{\sf R}-state by matrix elements of Eq.~(\ref{eq:QDM}) and string defects.  
Maroon diamonds denote results for configurations containing an
additional 8-link ``loop'' defect.     
All results are for a cubic 1024-bond cluster.
}
 \label{fig:Nf_ergodicity}
\end{figure}


We have repeated this analysis for more general local defects in the {\sf R}-state, 
seeding simulations from configurations generated using randomly chosen loops with 
10, 12, 14, 16, 18 and 20 links, applied states within the subsector containing 
the {\sf R}-state.  
After simulation starting from a configuration obtained using one 10-, 14- or 18-link 
loop we recover the same values of  $\langle m_{\sf R}(\mu) \rangle$ as in the maximally 
connected sector containing the {\sf R}-state. 
Meanwhile simulates started from a configuration obtained using an 12-, 16- or 20-link loop
exhibit the same values of $\langle m_{\sf R}(\mu) \rangle$ as those in the second 
large subsector accessed using the 8-link loop.  


This is consistent with our analysis of the transition graphs: the 8-loop cannot be undone by 
Hamiltonian matrix updates and so we remain in a subsector not connected to 
 {\sf R}-state, for which the average value of the order parameter has to be lower.
Loops of length 10, 14, 18... can typically be reduced to 6-link loops so that the resulting 
configuration remains in the maximally connected sector,  while loops of length 12, 16, 20... 
can be reduced to  an 8-link and 6-links loops, leading to the second large subsector. 
Although we do not have a precise understanding of the origin of the additional subsectors 
found in the 128-bond cluster [Table~\ref{sectors}], from our simulation for larger 
systems we tentatively conclude that the statistically accessible configurations with an 
even(odd) number of loops in their transition graph are of the type A and B 
(i.e. connected to the ${\sf R}$ state directly or by a single 8--link loop).
 

It is also interesting to compare the changes in the liquid phase ground state energy 
in detail.
The ground state energy in {\it all} subsectors must be zero at the RK point.
In the liquid phase bordering on the RK point, the gaps between states with different 
magnetic flux should scale as \mbox{$\Delta_\phi \sim \phi^2/L$} [Eq.~(\ref{eq:gaps})].
This in turn implies that, at the RK point, \mbox{$\Delta N_\phi \sim - \phi^2/L$} 
[Eq.~(\ref{eq:Nphi})].
In Fig.~\ref{fig:Nf_ergodicity} we present the results of simulations carried out
in maximally-connected subsector for each $\phi$, and in the subsector connected to 
it by a single, local, 8-link ``loop'' defect.  
The differences in the values of $\Delta N_\phi$ obtained in the two subsectors 
are smaller than the error bars on the simulation.   


This analysis represents only a partial solution to the problem of how to understand different 
subsectors within the Hilbert space of a three-dimensional QDM.  
However for the purposes of this paper, we gain confidence that GFMC simulations 
performed in the maximally-connected subsector, following the prescriptions of 
Section~\ref{model} and Section~\ref{simulation}, {\it do} give reliable 
estimates of the ground state properties of the QDM on a diamond lattice.


We can trust these estimates in the ordered, {\sf R}-state, because the alternative
subsectors correspond to excited states, and not to the ground state.
And we can trust the estimates in the quantum $U(1)$-liquid phase because in 
that case all statistically significant subsectors contribute equally to the ground state.
Thus, while the QDM on a diamond lattice is not ergodic within a given flux sector, 
this lack of ergodicity has no pathological consequences for the ground state phase 
diagram as we have calculated it.   
Our initial analysis suggests that the same may not be true of the QDM on a cubic 
lattice~\cite{olga-unpub}.   


\section{Conclusions}
\label{conclusions}


In this paper we set out to show that a three-dimensional quantum dimer model (QDM) --- 
the QDM on a bipartite diamond lattice --- could support a quantum liquid ground 
state, whose excitations include linearly dispersing photons and 
deconfined magnetic monopoles of an underlying $U(1)$ gauge theory.
We have identified the order parameter associated with the competing, ordered 
\mbox{``{\sf R}-state''} and, through use of quantum Monte Carlo simulation, have shown that 
it is finite only where the ratio $\mu$ of kinetic to potential energy terms in the QDM Hamiltonian 
is less than $\mu_c = 0.75 \pm 0.02$.
We have further shown that (a) the string-tension associated with separating a pair of 
monomers (magnetic monopoles) vanishes for $\mu > \mu_c$, and (b) the finite-size energy 
spectrum for $\mu_c < \mu \le 1$ is consistent with the predictions of the Maxwell action of 
conventional quantum electromagnetism.  


We conclude that the QDM on a diamond lattice undergoes a first-order phase transition
from an ordered {\sf R}-state into a quantum $U(1)$ liquid phase at $\mu = \mu_c$.
This liquid phase occupies an extended region of parameter space, terminating at the 
RK point $\mu = 1$.
This completes the analysis outlined in our earlier Letter [Ref.~\onlinecite{sikora09}], 
and confirms the validity of the phase diagram Fig.~\ref{fig:phase_diagram},  
previously proposed in Refs~\onlinecite{moessner03,bergman06-PRB} on 
the basis of field-theoretical arguments.


We have also explored how the Hilbert space of the QDM on a diamond lattice breaks up 
into different, disconnected, subsectors, and identified a new, Ising, quantum number 
associated with the number of loops in the transition graph of a given dimer configuration.
We argue that the lack of ergodicity within any given flux sector does not have pathological 
consequences for the ground state phase diagram, as found by quantum Monte Carlo 
simulations.
We further show that the same ground state phase diagram is obtained, regardless
of whether the dimers are treated as bosonic or fermionic objects.


Taken together, these results would seem to set the existence of a quantum $U(1)$ liquid 
phase in the QDM on a diamond lattice beyond reasonable doubt.
However the very fact that such a state exists in this model begs all manner of other 
questions.
Do other quantum dimer models on bipartite lattices in three dimension --- for example
the cubic lattice --- also support quantum $U(1)$ liquids, as conjectured in 
Ref.~\onlinecite{moessner03}~? 
What are the nature and consequences of the additional quantum numbers associated
with the non-ergodicity of three-dimensional quantum dimer models~?
Which other models can support a quantum $U(1)$ ground state, and for
what range of parameters ?   
An obvious candidate in this case is the quantum 
loop model studied in Ref.~\protect\onlinecite{hermele04}.  
Here we find evidence for a quantum $U(1)$ liquid for $-0.5 < \mu < 1$, which 
will be presented elsewhere~\cite{shannon-arXiv}.


The greatest excitement, however, would attend to finding a quantum $U(1)$ liquid 
phase in experiment, and identifying its deconfined, fractional excitations.
Here more quantitative analysis is needed, both to pin down the characteristics of these 
excitations, and to bridge the gap between idealized models such as Eq.~(\ref{eq:QDM}), 
and real materials or artificial assemblies of cold atoms.
Since the coherence temperature of the quantum $U(1)$ liquid will be low in either case,
both finite temperature effects~\cite{banerjee08}, and the consequences of competing 
interactions, need to be explored in more detail.  
%


\section{Acknowledgements}


It is our pleasure to acknowledge helpful conversations with 
Kedar Damle, 
Yong-Baek Kim, 
Andreas L\"auchli, 
Gregoire Misguich, 
Roderich Moessner, 
and Arno Ralko.   
We are particularly indebted to Federico Becca for advice and encouragement with 
simulation techniques.   
This work was supported under 
EPSRC grants EP/C539974/1 and EP/G031460/1, 
Hungarian OTKA grants K73455, 
U.S. National Science Foundation I2CAM International Materials Institute Award, Grant DMR-0645461, 
and the guest programs of MPI-PKS Dresden and YITP, Kyoto.


\appendix
\section{Absence of a fermionic sign problem}
\label{fermi}


All results for $H_{\text{QDM}}$ are obtained under the assumption that the dimers are bosonic 
objects, i.e., exchanging two dimers does not yield a phase. 
However, as we show in this Appendix, the same effective model can be used to describe 
the fermionic case. 


\begin{figure}[h]
\begin{center}
\includegraphics[width=5cm]{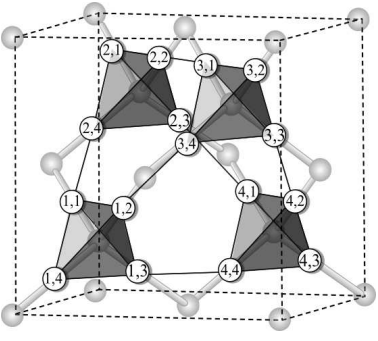}
\end{center}
\caption{
(Color online) 
Labeling of the sites of the pyrochlore lattice by enumerating one of the two types of tetrahedra 
and four sublattices on each tetrahedron. 
This labeling allows us to show that the Hamiltonian $H_{\text{QDM}}$  has no fermionic sign 
problem even if the underlying particles are fermions.
}
\label{fig:label}
\end{figure}


Let us now assume that the dimers are objects which are created (annihilated) by fermionic  
operators $c^{\dag}_{i,\sigma}$ ($c^{\vphantom{\dag}}_{i,\sigma}$).  
The index $i\in\{1,\dots, n\}$ refers to a tetrahedron and the index $\sigma\in\{1,2,3,4\}$ refers 
to the site on that tetrahedron (see Fig.~\ref{fig:label}).  
Since the diagonal part $\mathcal{N}_f$ of the  Hamiltonian in $H_{\text{QDM}}$  is not affected 
by changing to fermions, we only focus on the ring-exchange processes.  
These can be written in terms of the fermionic operators as 
\begin{equation}
{\mathcal K}_{\sf f}=\sum_{\smallhex}\left( 
c^{\dag}_{i_a,\sigma_a}c^{\vphantom{\dag}}_{i_a,\sigma_a'}
c^{\dag}_{i_b,\sigma_b}c^{\vphantom{\dag}}_{i_b,\sigma_b'}
c^{\dag}_{i_c,\sigma_c}c^{\vphantom{\dag}}_{i_c,\sigma_c'} +H.c.
\right),
\label{hamfer}
\end{equation}
where the sum is taken over all hexagons, and $i_a<i_b<i_c$.   
We thus choose only one of the twelve different ways of how the three fermions can hop 
around the hexagon. 


We show now that all twelve ways, which are resulting from clockwise/counter-clockwise 
processes as well as different sequences, have a positive matrix element. 
(i) Fermion configurations which result from clockwise or counter-clockwise processes are 
related by an even number of fermion exchanges, e.g. $(1,2,3)\leftrightarrow(2,3,1)$, and 
thus have the same sign. 
In passing we mention that this is generally true for ring-hopping processes which involve 
an \emph{odd} number of fermions, while processes with an \emph{even} number of fermions 
have the opposite sign and thus cancel each other.  
(ii) the matrix-element does not depend on the sequence in which the fermions hop.  
This can be seen if we choose ``ordered''  fermion configurations with respect to the 
tetrahedra, i.e., we fill the vacuum $|0\rangle$ by 
\begin{equation}
|c\rangle=c^{\dag}_{i_N,\sigma_N}\dots c^{\dag}_{i_2,\sigma_2}c^{\dag}_{i_1,\sigma_1}|0\rangle.
\end{equation}
Because of the hardcore-dimer constraint, there is at most one fermion on each tetrahedron 
$i$ which then has an arbitrary $\sigma$.  
In Hamiltonian (\ref{hamfer}), the fermionic operators come in pairs for each $i$ and thus 
there is always an even number of fermionic operators which need to be exchanged to 
evaluate the matrix elements. 
Thus, the sign does not depend on the sequence in which the operator pairs are applied 
and all non-zero matrix elements are equal to $+1$.  
The argument works only for the dimer model.  
If we consider a related loop model on the same lattice, there are two fermions per 
tetrahedron (i.e., two dimers attached to each site) and the sign problem can no 
longer be avoided using this method.


\section{Landau theory for {\sf R}-state}
\label{landau}


In this paper we studied the zero-temperature {\it quantum} phase
transition from the ordered \mbox{{\sf R}-state} into a quantum $U(1)$-liquid.  
However we note that the finite-temperature, phase transitions between ordered and 
classical $U(1)$ ``Coulomb'' phases in three-dimensional {\it classical} dimer models 
have also proved be a very interesting~\cite{huse03, alet06, powell08, chen09}.  
These phase transitions do not fall into the conventional Landau paradigm, 
since the high temperature Coulomb phase exhibits the power-law decay of correlation 
functions characteristic of a  $U(1)$ gauge theory, rather than the exponential decay 
predicted by a Ginzburg-Landau theory of the low-temperature ordered state.


In this context it is interesting to note that the present order parameter symmetry
permits a third-order invariant in a Landau action, as well as the usual 
quadratic and quartic terms~:
\begin{eqnarray}
\Delta_{\sf R} &=& a \sum_{i=1}^6 m_{{\sf R},i}^2 \nonumber \\ 
&&   +  b \left(
   m_{{\sf R},1} m_{{\sf R},2} m_{{\sf R},3} 
   + m_{{\sf R},4} m_{{\sf R},5} m_{{\sf R},6} \right) \nonumber \\ 
&&   + c_1 \sum_{i=1}^6 m_{{\sf R},i}^4 
   + c_2 \left(\sum_{i=1}^6 m_{{\sf R},i}^2 \right)^2 \nonumber \\ 
&& + c_3 \left[
  \left(m_{{\sf R},1}^2 +m_{{\sf R},2}^2 +m_{{\sf R},3}^2 \right)^2 \right. \nonumber \\ 
&&
+\left. \left(m_{{\sf R},4}^2 +m_{{\sf R},5}^2 +m_{{\sf R},6}^2 \right)^2 
  \right] \nonumber \\ 
&& 
+ c_4 \left(
    m_{{\sf R},1}^2 m_{{\sf R},4}^2 
  + m_{{\sf R},2}^2 m_{{\sf R},5}^2 
  + m_{{\sf R},3}^2 m_{{\sf R},6}^2 \right)^2  \nonumber \\ 
  &&
\label{Eq:invariant}
\end{eqnarray}
In the ordered phase with $\langle m_{\sf R} \rangle = m \ne 0$, the cubic term selects eight 
possible configurations.  
For $b<0$ these have the form given in Table~\ref{table:cubic}, and are of course the 
eight \mbox{{\sf R}-state} configurations, which become exact ground states of the quantum 
problem in the limit $\mu \to -\infty$. 
For $b>0$ the cubic term does not select a valid (static) dimer configuration.


\begin{table}
\caption{
Eight \mbox{{\sf R}-states} selected by the cubic invariant $b<0$ in Eq.~(\ref{Eq:invariant}), 
expressed in terms of the coefficients of the order parameter Eq.~(\ref{eq:op}), 
within an ordered state with $m =\langle m_{\sf R}/\sqrt{3} \rangle$.  
The four \mbox{{\sf R}-states} labelled $\zeta=1,\ldots,4$ have the opposite chirality from those 
labelled $\zeta=5,\ldots,8$.
}  
\begin{ruledtabular}
\begin{tabular}{|c|c|c|c|c|c|c|}
\label{table:cubic}
$\zeta$ &$m_{{\sf R},1}$ & $m_{{\sf R},2}$ & $m_{{\sf R},3}$ & 
   $m_{{\sf R},4}$ & $m_{{\sf R},5}$ & $m_{{\sf R},6}$ \\
\hline
1& $m$ & $m$ & $m$ & 0 & 0 & 0  \\ 
2 & $m$ & $m$ & $-m$ & 0 & 0 &0 \\ 
3 & $-m$ & $m$ & $-m$ & 0 & 0 & 0 \\ 
4 & $-m$ & $-m$ & $m$ & 0 & 0 & 0  \\ 
\hline
5 & 0 & 0 & 0 & $m$ & $m$ & $m$ \\ 
6 & 0 & 0 & 0 & $m$ & $-m$ & $-m$ \\ 
7 & 0 & 0 & 0 & $-m$ & $m$ & $-m$  \\ 
8 & 0 & 0 & 0 & $-m$ & $-m$ & $m$ 
\end{tabular}
\end{ruledtabular}
\end{table}


\section{Comparison of variational wave functions}
\label{VMC}


In this Appendix we explore the transition from solid to liquid phases of the quantum 
dimer model on a diamond lattice using a variety of variational wave functions.
This is an essential preliminary to Green's function Monte Carlo (GFMC) simulations, 
which rely on importance sampling of configurations to achieve convergence in a finite time.
Except where stated otherwise, the calculations presented here were performed using 
variational Monte Carlo, within the stochastic reconfiguration 
algorithm~\cite{sorella01,capello05}, for a 1024-bond cluster with cubic symmetry.  


\begin{figure}[h]
\begin{center}
\includegraphics[width=8cm]{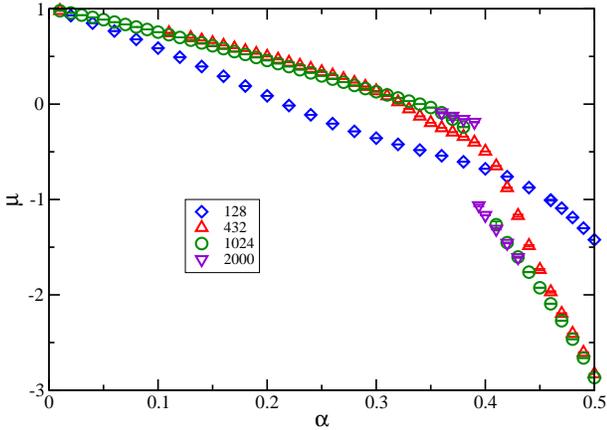}
\end{center}
\caption{
(Color online) 
Value of $\mu$ for which the one-parameter variational wave function 
$|\psi^{\sf Var}_\alpha \rangle$ [cf. Eq.~(\ref{eq:Nflip})] has the lowest energy, 
as a function of the variational parameter $\alpha$.
Results are plotted for cubic clusters ranging in size from 128 to 2000 bonds.
}
\label{fig:mu_vs_alpha}
\end{figure}


Perhaps the simplest wave function we can consider is
\begin{eqnarray}
|\psi^{\sf Var}_\alpha \rangle_{\phi, {\sf c}}  
= \exp\left[\alpha {\mathcal N}_{\sf f} \right] | \psi_{\sf RK} \rangle_{\phi, {\sf c}} 
\label{eq:Nflip}
\end{eqnarray}
where $\alpha$ is the single variational parameter associated with the total number of 
flippable hexagons ${\mathcal N}_{\sf f}$, and $|\psi_{\sf RK}\rangle_{\phi, {\sf c}}$ is the 
RK point ground state function, i.e. the equally weighted sum of all dimer configurations within 
a given subsector of the Hilbert space.
In the discussion of ground state properties which follows, we will consider the maximally 
connected subsector for $\vec\phi = 0$, which comprises all dimer configurations connected
to an {\sf R}-state by matrix elements of Eq.~(\ref{eq:QDM}).  


We can view Eq.~(\ref{eq:Nflip}), as the Jastrow form which follows from Eq.~(\ref{eq:pert}).   
For this reason we anticipate that this wave function will work well in the vicinity 
of the RK point, but will tend to over-emphasize any liquid phase occurring for $\mu \to 1$.   
The wave function $|\psi^{\sf Var}_\alpha \rangle_{\phi, {\sf c}}$ has the advantage that the 
value of $\mu$ which minimizes the ground state energy for a given $\alpha$ can be found by 
evaluating averages at the RK point
\begin{eqnarray}
 \mu  
  =\frac{\frac{1}{2}
  \langle \mathcal{N}_{\sf f} \mathcal{K}_{\sf f} 
  + \mathcal{K}_{\sf f} \mathcal{N}_{\sf f} \rangle_{\sf RK} 
  - \langle \mathcal{K}_{\sf f} \rangle_{\sf RK} 
  \langle  \mathcal{N}_{\sf f} \rangle_{\sf RK} 
  }
  {\langle \mathcal{N}_{\sf f}^2  \rangle_{\sf RK} 
-  \langle \mathcal{N}_{\sf f} \rangle_{\sf RK}^2 }
\label{eq:karlo_n}
\end{eqnarray}
where $\mathcal{K}_{\sf f}$ is defined by Eq.~(\ref{eq:K}).   
In Fig.~\ref{fig:mu_vs_alpha} we present results for $\mu$ as a
function of $\alpha$ for cubic clusters ranging in size from 128 to 1024 
diamond lattice bonds.  
It is tempting to interpret the abrupt change in $\partial \mu/\partial \alpha$ 
at $\alpha \approx 0.4$, $\mu \approx -0.5$ as evidence of 
a first order phase transition.   
However comparison with exact diagonalization, and with other variational
wave functions below, suggests that $|\psi^{\sf Var}_\alpha \rangle$ gives 
very poor energies away from the RK point, and is only really useful for $\mu \to 1$.
 

\begin{figure}[h]
\begin{center}
\includegraphics[width=8cm]{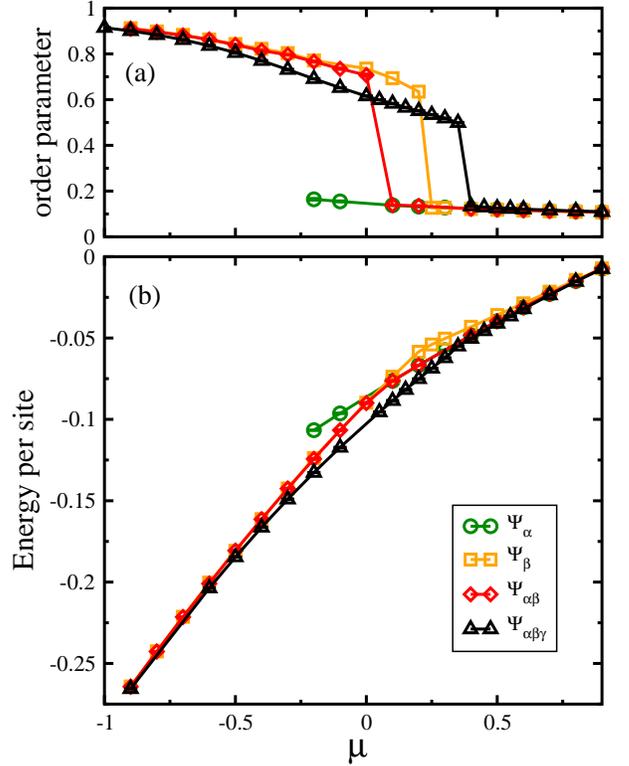}
\end{center}
\caption{
(Color online) 
Comparison of different variational wave functions for 1024-bond cluster: 
(a) order parameter and (b) energy per site: 
(green) $|\psi^{\sf Var}_\alpha \rangle$ [cf.~Eq.~(\ref{eq:Nflip})] 
(orange) $|\psi^{\sf Var}_\beta \rangle$ [cf.~Eq.~(\ref{eq:beta})]; 
(red) $|\psi^{\sf Var}_{\alpha\beta} \rangle$[cf.~Eq.~(\ref{eq:alpha_beta}]); 
(black) $|\psi^{\sf Var}_{\alpha\beta\gamma} \rangle$ [cf.~Eq.~(\ref{eq:alpha_beta_gamma})]. 
}
\label{fig:vmc_energy_OP}
\end{figure}


\begin{figure}[h]
\begin{center}
\includegraphics[width=8cm]{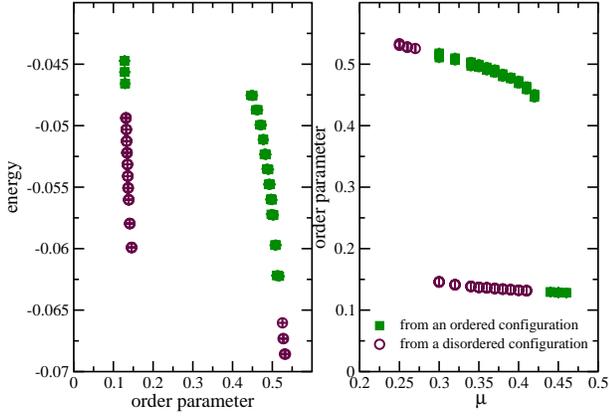}
\end{center}
\caption{
(Color online) 
Quantum hysteresis --- variational Monte Carlo results for 
$|\psi^{\sf Var}_{\alpha\beta\gamma} \rangle$ 
[cf.~Eq.~(\ref{eq:alpha_beta_gamma})] in a 1024-bond cluster.  
(left) energy per site as a function of the order parameter $m_{\sf R}$, 
(right) order parameter as a function of $\mu$.  
Simulations converge to different local minima for a range of values of 
$0.28 < \mu <  0.43$, according to whether they are started from an ordered 
or a disordered state.
}
\label{Fig:histeresis_1024}
\end{figure}


We can construct another simple one-parameter variational wave function based
on the order parameter $m_{\sf R}$ for the {\sf R}-state
\begin{eqnarray}
| \psi^{\sf Var}_\beta \rangle_{\phi, {\sf c}} =\exp[\beta m_{\sf R}] |\psi_{\text{RK}}\rangle_{\phi, {\sf c}} 
\label{eq:beta}
\end{eqnarray}
This wave function gives a good account of the ordered {\sf R}-state, 
and exhibits a clear first order phase transition transition from 
as a jump in $\langle m_{\sf R}\rangle$ for $\mu \approx 0.25$  in a 1024-bond
cubic cluster --- cf. Fig.~\ref{fig:vmc_energy_OP}(a).
However from Fig.~\ref{fig:vmc_energy_OP}(b) we can see that $| \psi^{\sf Var}_\alpha \rangle$
has a lower energy approaching the RK point.


Seeking the advantages of both terms, we try a two-parameter wave function 
\begin{eqnarray}
|\psi^{\sf Var}_{\alpha\beta}\rangle_{\phi, {\sf c}}
=\exp[\alpha {\mathcal N}_{\sf f}  + \beta m_{\sf R}] |\psi_{\text{RK}}\rangle_{\phi, {\sf c}}
\label{eq:alpha_beta}
\end{eqnarray}
This gives lower energies overall, and predicts a transition from solid to liquid 
for $\mu \approx 0.2$  in a (1024 bonds).  
However this wave function does a poor job of describing the quantum melting 
of the {\sf R}-state.


In order to better model this, we introduce correlations between individual flippable 
plaquettes, on which all dynamics in the QDM depend
\begin{eqnarray}
|\psi^{\sf Var}_{\alpha\beta\gamma} \rangle_{\phi, {\sf c}} 
= \exp[\alpha {\mathcal N}_{\sf f} + \beta  m_{\sf R} 
+ \sum_{ij} \gamma_{ij} \tau_i \tau_j ] |\psi_{\text{RK}}\rangle_{\phi, {\sf c}}
\nonumber\\
\end{eqnarray}
Here $\tau_i =1$ if the plaquette $i$ is flippable and $\tau_i =0$, otherwise.
For complete generality we should also allow for the chirality of the flippable
plaquette, and consider a separate $\gamma_{ij}$ for each 
indistinguishable pair of plaquettes in the cluster.  
For the smallest 128~bond cluster there are 16 of these, and the number grows 
to 84 in the 1024-bond cluster.  
However in practice we do not find that including chirality brings 
significant gains in energy, and we restrict our calculations to a set of 
about 40 $\gamma_{ij}$.


This wave function has a significantly lower energy than any of the alternatives considered
above for $-0.5 < \mu < 0.25$, and predicts a transition for $\mu \approx 0.4$ 
(1024-bond cluster --- black lines in Fig.~\ref{fig:vmc_energy_OP}).   
The wave function also treats fluctuations accurately enough to capture some 
a part of the hysteresis associated the quantum melting of a solid.
In Fig.~\ref{Fig:histeresis_1024} we show the order parameter and energy for the 
1024-bond cluster, obtained using $|\psi^{\sf Var}_{\alpha\beta\gamma} \rangle$, 
seeding VMC from different states.  
Starting from an ordered state, we obtain an ordered state solution for $\mu \lesssim 0.42$, 
while starting from a disordered configuration we could see a minimum for $\mu \gtrsim 0.3$.
These results confirm the first order character of the phase transition, and 
extend the usefulness of $|\psi^{\sf Var}_{\alpha\beta\gamma} \rangle$ as a guide
function for GFMC some way into the region where it predicts a liquid phase.


\section{Series expansion about the ordered {\sf R}-state}
\label{perturbation}


\begin{figure}[tbp]
\begin{center}
\includegraphics[width=8cm]{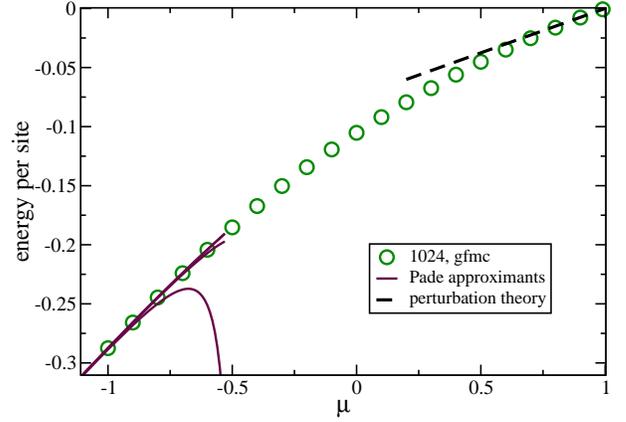}
\end{center}
\caption{(Color online) 
Energy per site as a function of $\mu$. GFMC results for 
1024-bond cluster (points) are compared to Pade approximants for series expansion 
around an ordered state (solid lines) and perturbation expansion in the vicinity of the 
RK point (dashed lines).}
 \label{Fig: pade_gsen}
\end{figure}


\begin{figure}[tbp]
\begin{center}
\includegraphics[width=8cm]{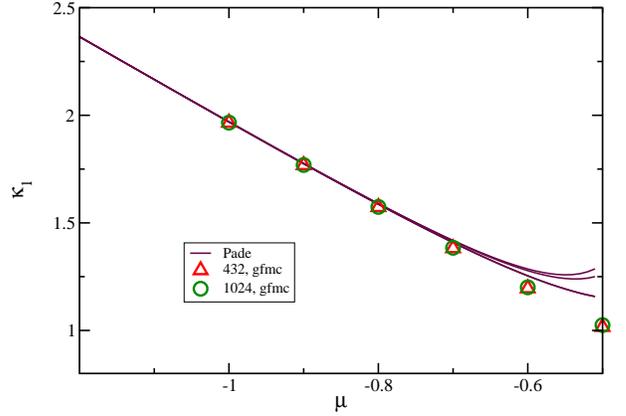}
\end{center}
\caption{(Color online) 
Monopole string tension $\kappa_1 = \Delta_1/L$, plotted for a range of $\mu$ spanning 
the ordered \mbox{{\sf R}-state} phase.
Solid lines denote different Pad\'e approximants to the series expansion 
about an \mbox{{\sf R}-state} configuration [cf. Eq.~(\ref{eq:string_tension_series})].  
Points denote the results of GFMC simulations of 432-bond (red) and 
1024-bond (green) cubic clusters. 
}
 \label{Fig: pade_gap}
\end{figure}


There are two limits in which the exact ground state of Eq.~(\ref{eq:QDM}) is known.  
One is the RK point $\mu=1$, where the ground state is the equally weighted sum over all 
dimer configurations.
The other is $\mu\to-\infty$, where the ground state is one of 
eight degenerate {\sf R}-state configurations.  
In the second case we can construct a perturbation theory in $1/\mu$ by expanding 
fluctuations about an ordered {\sf R}-state.  
We find that the ground state energy (per bond) in the 
zero-flux sector behaves as
\begin{eqnarray}
\frac{E_0}{N} &=& \frac{\mu}{4} +\frac{1}{24}\frac{1}{\mu} -\frac{229} {47520}\frac{1}{\mu^3}   \nonumber\\
   && + \frac{37811909} {48926592000}\frac{1}{\mu^5} + {\mathcal O}\left(\frac{1}{\mu^7} \right) \;.
\label{eq:seriesE0}
\end{eqnarray}
The first term in this expansion counts the number of flippable plaquettes in the 
\mbox{{\sf R}-state} (a quarter of the number of bonds).   
Subsequent terms are derived combinatorially, by counting the number
of possible ways of returning to the the initial {\sf R}-state, using a given 
number of applications of ${\mathcal K}_{\sf f}$ and following the recipe of the 
Rayleigh-Sch\"odinger perturbation theory.   
This can be calculated for a finite-size cluster, or an infinite lattice;  
the results quoted here are for an infinite lattice.  
The resulting series Eq.~(\ref{eq:seriesE0}) contains only terms 
of odd order in $1/\mu$.


The series for $E_0/N$ [Eq.~(\ref{eq:seriesE0})] can be continued to 
values of $\mu$ relevant to our simulations by means of  Pad\'e approximants.  
In Fig.~\ref{Fig: pade_gsen} we compare several different Pad\'e approximants to 
Eq.~(\ref{eq:seriesE0}), with the ground state energy per bond found in GFMC 
simulations of a 1024-bond cluster.   
We find essentially perfect agreement between our simulations results and the results
of perturbation theory for $\mu \lesssim -0.8$, at which point the different 
Pad\'e approximants start to diverge.    
Some of the Pad\'e approximants remain in agreement with the simulations up to 
$\mu \approx -0.5$.   
For comparison, we also include results for $E_0/N$ derived in an expansion
about the RK point, as described in Section~\ref{liquid}.   
Again, the agreement between perturbation theory and simulation is found to be
very good in the limit  $\mu \to 1$.


For $\mu \to -\infty$, it is also possible to perform a series about a state with unit-flux, i.e.
an {\sf R}-state with a single string excitation. From Eq.~(\ref{eq:string_tension}) we find 
the monopole string tension to be 
\begin{eqnarray}
\kappa_1 &=& 
-2\mu + \frac{1}{15}\frac{1}{\mu} -\frac{71243}{1995840}\frac{1}{\mu^{3}}
\nonumber\\&&
+\frac{1044561213049}{41535007113600000}\frac{1}{\mu^{5}}
+ {\mathcal O}\left(\mu^{-7}\right) \;.
\label{eq:string_tension_series}
\end{eqnarray}
In Fig.~\ref{Fig: pade_gap} we compare the perturbational expansion and GFMC simulation 
results for the string tension $\kappa_1$.
The prediction of the perturbation theory for $\mu \alt -1$ matches 
well to the results of simulations for $\mu \agt -1$.


\bibliographystyle{apsrev}


\end{document}